\documentclass[12pt]{article}

\usepackage{xspace,colortbl}
\usepackage{graphicx}
\usepackage{amsmath}
\usepackage{amsthm}
\usepackage{amssymb}
\usepackage{latexsym}
\usepackage{indentfirst}
\usepackage{setspace}
\usepackage[top=1.2in,bottom=1in]{geometry}

\begin{document}

\title{\Large\bf On Constructing the Asymptotic Solutions for Phase Transitions in a Slender Cylinder Composed of a Compressible Hyperelastic Material with Clamped End Conditions}
\author{\small Hui-Hui Dai$^1$, Jiong Wang$^2$ and Zhen Chen$^3$
\\\small $^1$Department of Mathematics and Liu Bie Ju Centre for Mathematical Sciences,\\\small City University of Hong Kong,\ 83 Tat Chee Avenue, Kowloon Tong, Hong
Kong\\ \small Email: mahhdai@cityu.edu.hk
\\\small $^2$Department of
Mathematics, City University of Hong Kong, 83 Tat Chee
Avenue,\\\small Kowloon Tong, Hong Kong\\ \small Email:
jiongwang2@student.cityu.edu.hk
\\\small $^3$Department of Civil and Environmental Engineering,\\ \small  University of Missouri-Columbia,Columbia, MO 65211-2200, USA
}

\date{}
\maketitle

%%%%%%%%%%%%%%%%%%%%%%%%%%%%%%%%%%%%%%%%%%%%%%%%%%%%%%%%%%%%%%%%%%%%%%%%%%%%%%%%%%%%%%%%%%%%%%%%

\begin{abstract}
In this paper, we study phase transitions in a slender circular
cylinder composed of a compressible hyperelastic material with a
non-convex strain energy function. We aim to construct the
asymptotic solutions based on an axisymmetrical three-dimensional
setting and use the results to describe the key features (in
particular, instability phenomena) observed in the experiments by
others. The difficult problem of the solution bifurcations of the
governing nonlinear partial differential equations (PDE's) is solved
through a novel approach. By using a methodology involving coupled
series-asymptotic expansions, we derive the normal form equation of
the original complicated system of nonlinear PDE's. By writing the
normal form equation into a first-order dynamical system and with a
phase-plane analysis, we manage to deduce the global bifurcation
properties and to solve the boundary-value problem analytically. The
asymptotic solutions (including post-bifurcation solutions) in terms
of integrals are obtained. The engineering stress-strain curve
plotted from the asymptotic solutions can capture the key features
of the curve measured in a few experiments (e.g., the stress drop,
the stress plateau, and the small stress valley). It appears that
the asymptotic solutions obtained shed certain light on the
instability phenomena associated with phase transitions in a
cylinder, in particular the role played by the radius-length ratio.
Also, an important feature of this work is that we consider the
clamped end conditions, which are more practical but rarely used in
literature for phase transition problems.
\end{abstract}

\noindent\textbf{Key words}: Phase transformation; Hyperelastic
material; Asymptotic analysis; Cylinder; Bifurcations of PDE's

%%%%%%%%%%%%%%%%%%%%%%%%%%%%%%%%%%%%%%%%%%%%%%%%%%%%%%%%%%%%%%%%%%%%%%%%%%%%%%%%%%%%%%%%%%%%%%%%

\section{Introduction}

Applications of phase-transforming materials such as shape memory
alloys (SMAs) and shape memory polymers are very broad. For example,
they have been used to make satellite dampers, golf club heads and
snake-like robots and so on. In particular, these materials have
been used to design many minimal surgery devices (Pelton \emph{et
al.} 1997). A deep and thorough understanding of the behaviour of
this type of phase-transforming material is essential in the
manufacturing and designing of these devices.

Systematic experiments on uniaxial extensions of superelastic NiTi
alloy (a kind of shape memory alloys and also one kind of
phase-transforming materials) wires and strips (Shaw $\&$ Kyriakides
1995, 1997, 1998) showed the measured engineering stress-strain
curves have the key features: The nucleation stress occurs at a
local maximum which is significantly larger than the Maxwell stress;
following the nucleation stress there is a sharp stress drop; and
afterwards there is a stress plateau. These features were also
observed in experiments done by others (Sun \emph{et al.} 2000, Tse
$\&$ Sun 2000, Favier \emph{et al.} 2001 and Li $\&$ Sun 2002).

Theoretically, solid-solid phase transitions have also been studied
for a long time in the context of both continuum and lattice
theories. The seminal work of Ericksen (1975), which considered a
continuum one-dimensional stress problem, made clear that for a
non-convex strain energy function the solution with two phases can
arise and there are multiple solutions. Based on the lattice model
for a two-phase martensitic material, it is possible to deduce in
the related continuum model the strain-energy function has double
wells; see Ball $\&$ James (1992). In general, it is now understood
that in the continuum scale for a material whose strain-energy
function is non-convex phase transitions can take place (see also,
Abeyaratne \emph{et al.} 2001). With a proper choice of a
strain-energy function, phase transitions can be modeled through a
continuum theory. Still, to justify this point of view, it is
desirable (probably necessary) to compare the analytical solutions
based on this type of energy functions with experimental results.
But, the difficulty is that we lack mathematical theories for mixed
type equations that typically arise from non-convex energy
functions. Analytical solutions for boundary-value problems are very
few. In the classical paper of Ericksen (1975), analytical solutions
were constructed for a static problem based on a pure
one-dimensional stress model, which neglects the effect from other
dimensions. However, a one-dimensional model appears to be not
sophisticated enough to capture some key features observed in
experiments. We give the explanations below.

In the experiments (Shaw $\&$ Kyriakides 1995, 1997; Li $\&$ Sun
2002), it was observed that after the nucleation stress was reached
the nucleation process began and accompanying with it there was a
radial contraction (necking). Also, after the two-phase state was
formed, the deformation was inhomogeneous, one part being thin and
one part being thick. As pointed out by Chang \emph{et al.} (2006)
that the axial extent of the transformation front is of the order of
the radius. Thus, to model the nucleation process and the
inhomogeneous deformation of different thicknesses, one should
consider the radial deformation.

Theoretically, if one treats the problem as a one-dimensional stress
one, there is a discontinuous interface between two phases with two
different strain values. The thicknesses of the cylinder at two
phases are different, thus the shear strain $U_Z$ ($U$ is the radial
displacement and $Z$ is the axial coordinate) is nonzero (infinite
at the phase boundary). The traction-free boundary conditions
require the two stress components $\Sigma_{Rr}$ (depending on
$U_Z^2$ up to the second order) and $\Sigma_{Rz}$ (depending on
$U_Z$) to be zero at the lateral surface, which cannot be satisfied.
Thus, it appears that to model phase transitions by a
one-dimensional stress approximation violates the traction-free
boundary conditions.

In the papers of Dai $\&$ Cai (2006) and Cai $\&$ Dai (2006), phase
transitions in a slender cylinder composed of a \emph{special}
\emph{incompressible} elastic material was considered. A novel
series-asymptotic approach is utilized to reduce the field
equations. A proper asymptotic model equation is derived, which
takes into account the influences of the radial deformation and
traction-free boundary conditions. The solutions for two
boundary-value problems are obtained, and they could capture the key
features observed in experiments.

In this paper, we study the phase transitions in a slender cylinder
composed of a \emph{general} \emph{compressible} elastic material
due to tension/extension. Also, different from the previous studies
on \emph{incompressible} materials, here we further consider the
coupling effect of the material nonlinearity and geometrical size.
Another important new feature is that we consider the more practical
clamped end conditions instead of the natural boundary conditions
used in Cai $\&$ Dai (2006).

We consider the problem in a three-dimensional setting, different
from the one-dimensional stress problem studied by Ericksen (1975).
However, the strain-energy function is assumed to have the same
property as that in Ericksen (1975), i.e., for a one-dimensional
stress problem the stress-strain curve has a peak-valley combination
(cf. figure 1). We aim at constructing the asymptotic solutions and
using them to explain the experimental results.

Mathematically, to deduce the analytical solutions for the present
problem is a very difficult task. One need to deal with coupled
nonlinear partial differential equations (PDE's) together with
complicated boundary conditions. Further, the existence of multiple
solutions (corresponding to the instability phenomena (e.g., stress
drop) observed in experiments) makes the problem even harder to
solve. Here, the analysis is carried out by a novel methodology
developed earlier (Dai $\&$ Huo 2002; Dai $\&$ Fan 2004; Dai $\&$
Cai 2006), which is capable of treating the global bifurcation
problem of nonlinear PDE's and obtaining the post-bifurcation
solutions. We construct the solutions and extract from them
important information on the deformation configurations, the
nucleation stress, the instability phenomena and the transformation
front. Comparisons with experimental results are made, which show
that the asymptotic solutions can capture the key features of the
experimental engineering stress-strain curves and the instability
phenomena as observed in experiments. The qualitative agreements
give supporting evidence that a non-convex strain energy function
can be used to describe solid-solid phase transitions and the
instability phenomena in phase transformations are mainly due to the
non-convexity of the strain energy function.

This paper is arranged as follows. In section 2, we formulate the
field equations by treating the slender cylinder as a
three-dimensional object. In section 3, we carry out a
non-dimensionalization process to extract the important small
variable and two small parameters which characterize this problem.
Then we derive the normal form equation of the original governing
nonlinear PDE's in section 4, through series and asymptotic
expansions. In section 5, we show that the Euler-Lagrange equation
can also lead to the same normal form equation, which justifies our
method in deriving this equation. In section 6, we propose the
clamped boundary conditions. In section 7, we construct the
asymptotic solutions for both a force-controlled problem and a
displacement-controlled problem. We also use the solutions obtained
to explain some experimental results. Finally, some conclusions are
drawn.

%%%%%%%%%%%%%%%%%%%%%%%%%%%%%%%%%%%%%%%%%%%%%%%%%%%%%%%%%%%%%%%%%%%%%%%%%%%%%%%%%%%%%%%%%%%%%%%%

\section{Three-dimensional field equations}

We consider the axisymmetric deformations of a slender elastic
cylinder subject to a static axial force at two ends. The lateral
surface is traction-free and the end conditions will be considered
later. In an undeformed state, the radius of the cylinder is $a$ and
the total length is $2l$. It is assumed that $\delta=a/l<<1$. We
take the cylindrical polar coordinate system and denote $(R, \Theta,
Z)$ and $(r, \theta, z)$ the coordinates of a material point of the
cylinder in the reference and current configurations, respectively.
The finite radial and axial displacements can be written as
$$
U(R,Z)=r(R,Z)-R,\ \ \ W(R,Z)=z(R,Z)-Z.\ \ \eqno(2.1)
$$
We introduce the orthonormal bases associated with the cylindrical
coordinates and denote these by $\textbf{E}_R$, $\textbf{E}_\Theta$,
$\textbf{E}_Z$ and $\textbf{e}_r$, $\textbf{e}_\theta$,
$\textbf{e}_z$ in the reference and current configurations,
respectively. Then the deformation gradient tensor $\textbf{F}$ is
given by
$$
\textbf{F}=(1+U_R)\textbf{e}_r\otimes\textbf{E}_R+U_Z
\textbf{e}_r\otimes \textbf{E}_Z+(1+\frac{U}{R}) \textbf{e}_\theta
\otimes \textbf{E}_\Theta+W_R \textbf{e}_z \otimes
\textbf{E}_R+(1+W_Z) \textbf{e}_z \otimes \textbf{E}_Z.\ \
\eqno(2.2)
$$
For an hyperelastic material, the strain energy function $\Phi$ is a
function of the three invariants $I_1$, $I_2$ and $I_3$ of the left
Cauchy-Green strain tensor $\textbf{B}=\textbf{F}\textbf{F}^T$; that
is, $\Phi=\Phi(I_1,I_2,I_3)$. We suppose that $\Phi$ is non-convex
in a pure one-dimensional stress problem such that phase transition
can take place. The nominal stress tensor $\bf{\Sigma}$ is given by
$$
\textbf{$\Sigma$}=\frac{\partial \Phi}{\partial \textbf{F}},\ \ \
\Sigma_{ji}=\frac{\partial\Phi}{\partial F_{ij}}. \eqno(2.3)
$$

If the strains are small, it is possible to expand the nominal
stress components in term of the strains up to any order. The
formula containing terms up to the third order material nonlinearity
is (cf. Fu $\&$ Ogden 1999)
$$
\Sigma_{ji}=a_{jilk}^1\eta_{kl}+\frac{1}{2}a^2_{jilknm}\eta_{kl}\eta_{mn}+\frac{1}{6}a^3_{jilknmqp}\eta_{kl}\eta_{mn}\eta_{pq}+O(|\eta_{st}|^4),\
\ \eqno(2.4)
$$
where $\eta_{ij}$ is the components of the tensor
$\textbf{F}-\textbf{I}$, $a_{jilk}^1$, $a^2_{jilknm}$ and
$a^3_{jilknmqp}$ are incremental elastic moduli, which can be
calculated once a specific form of the strain-energy function is
given, and the formulas can be found in Appendix A. From the formula
(2.4), we could obtain the nominal stress components $\Sigma_{ji}$.
For example,
$$
\begin{aligned}
\Sigma_{Zz}=&\xi_2\frac{U}{R}+\xi_1W_Z+\xi_2U_R+\frac{1}{2}(\eta_2\frac{U^2}{R^2}+\eta_4U_Z^2+2\eta_2\frac{UW_Z}{R}+\eta_1W_Z^2+2\eta_3\frac{UU_R}{R}\\
&+2\eta_2W_ZU_R+\eta_2U_R^2+2\eta_7U_ZW_R+\eta_4W_Z^2)+\frac{1}{6}(\theta_2
\frac{U^3}{R^3}+3\theta_8\frac{UU_Z^2}{R}\\
&+3\theta_3\frac{U^2W_Z}{R^2}+3\theta_5U_Z^2W_Z+3\theta_2\frac{UW_Z^2}{R}+\theta_1W_Z^3+3\theta_4\frac{U^2U_R}{R^2}+3\theta_7U_Z^2U_R\\&
+6\theta_4\frac{UU_RW_Z}{R}+3\theta_2U_RW_Z^2+3\theta_4\frac{UU_R^2}{R}+3\theta_3U_R^2W_Z+\theta_2U_R^3+6\theta_{15}\frac{UU_ZW_R}{R}\\
&+6\theta_{12}U_ZW_ZW_R+6\theta_{14}U_RU_ZW_R+3\theta_8\frac{UW_R^2}{R}+3\theta_5W_R^2W_Z+3\theta_7U_RW_R^2),
\end{aligned}
\eqno(2.5)
$$
where $\xi_i$, $\eta_j$ and $\theta_k$ are some elastic moduli,
whose formulas are given in Appendix A. The other non-zero stress
components can also be obtained but we omit their lengthy
expressions for brevity. Owing to the complexity of calculations, we
shall only work up to the third-order material nonlinearity. In the
experiments, the maximum strain is less than $10\%$, and such an
approximation is sufficiently accurate.

$\bf{\Sigma}$ satisfies the following field equations:
$$
\textrm{Div}(\Sigma)=0, \ \ \eqno(2.6)
$$
which yields the following equations
$$
\frac{\partial\Sigma_{Zz}}{\partial Z}+\frac{\partial
\Sigma_{Rz}}{\partial R}+\frac{\Sigma_{Rz}}{R}=0,\ \ \eqno(2.7)
$$
$$
\frac{\partial \Sigma_{Rr}}{\partial R}+\frac{\partial
\Sigma_{Zr}}{\partial
Z}+\frac{\Sigma_{Rr}-\Sigma_{\Theta\theta}}{R}=0.\ \ \eqno(2.8)
$$
We consider the case that the lateral surface of the cylinder is
traction-free. Thus, we have the boundary conditions
$$
\Sigma_{Rr}|_{R=a}=0,\ \ \ \ \Sigma_{Rz}|_{R=a}=0.\ \ \eqno(2.9)
$$
We shall derive the asymptotic solutions of (2.7) and (2.8) under
(2.9) and some end conditions for a given external force.

Equations (2.7) and (2.8) together with (2.4) provide the governing
equations for two unknowns $U$ and $W$. The formal two are very
complicated nonlinear partial differential equations (PDE's) and the
boundary conditions (2.9) are also complicated nonlinear relations
(cf. (3.3)-(3.6)). To describe the instability phenomena during
phase transitions, one need to study the bifurcations of this
complicated system of nonlinear PDE's. Here, we shall adopt an
alternative approach involving coupled series-asymptotic expansions
to tackle this bifurcation problem. A similar methodology has been
developed to study nonlinear waves and phase transitions in
incompressible materials (see Dai $\&$ Huo 2002, Dai $\&$ Fan 2004,
Dai $\&$ Cai 2006, Cai $\&$ Dai 2006). First, we shall
nondimensionalize this system to identify the relevant small
variable and small parameters.

%%%%%%%%%%%%%%%%%%%%%%%%%%%%%%%%%%%%%%%%%%%%%%%%%%%%%%%%%%%%%%%%%%%%%%%%%%%%%%%%%%%%%%%%%%%%%%%%%%%%%

\section{Non-dimensional equations}

We first introduce a very important transformation
$$
U=vR,\ \ \ s=R^2,\ \eqno(3.1)
$$
then do the following scalings:
$$
s=l^2\tilde{s},\ \ \ Z=l\tilde{z},\ \ \ W=h\tilde{w},\ \ \
v=\frac{h}{l}\tilde{v},\ \ \ \epsilon=\frac{h}{l},\ \ \ \eqno(3.2)
$$
where $l$ is the length of the cylinder, $h$ is a characteristic
axial displacement, and $\epsilon$ is regarded to be a small
parameter (equivalent to a small engineering strain). Substituting
(3.1) and (3.2) into (2.7) and (2.8), we obtain
$$
2(\xi_2+\xi_3)v_z+\xi_1w_{zz}+4\xi_3w_s+s[2(\xi_2+\xi_3)v_{sz}+4\xi_3w_{ss}]+\cdots=0,\
\ \eqno(3.3)
$$
$$
\xi_3v_{zz}+8\xi_1v_s+2(\xi_2+\xi_3)w_{sz}+s4\xi_1v_{ss}+\cdots=0.\
\ \ \eqno(3.4)
$$
Here and hereafter,we have dropped the tilde for convenience. The
full forms of (3.3) and (3.4) are very lengthy and can be found in
Appendix B, here we just present the first few terms. Substituting
(3.1) and (3.2) into the traction-free boundary conditions (2.9), we
obtain
$$
\begin{aligned}
&(\xi_1+\xi_2)v+\xi_2w_z+s2\xi_1v_s+\epsilon[(\frac{1}{2}\eta_1+\frac{3}{2}\eta_2)v^2+(\eta_2+\eta_3)vw_z+\frac{1}{2}\eta_2w_z^2
\\&+s(\frac{1}{2}\eta_4v_z^2+(2\eta_1+2\eta_2)vv_s+2\eta_2w_zv_s+2\eta_7v_zw_s+2\eta_4w_s^2)+s^22\eta_1v_s^2]
\\&+\epsilon^2[(\frac{1}{6}\theta_1+\frac{2}{3}\theta_2+\frac{1}{2}\theta_3)v^3+(\frac{1}{2}\theta_2+\frac{3}{2}\theta_4)v^2w_z+(\frac{1}{2}\theta_3+\frac{1}{2}\theta_4)vw_z^2+\frac{1}{6}\theta_2w_z^3
\\&+s((\frac{1}{2}\theta_5+\frac{1}{2}\theta_8)vv_z^2+\frac{1}{2}\theta_7v_z^2w_z+(\theta_1+2\theta_2+\theta_3)v^2v_s+(2\theta_2+2\theta_4)vw_zv_s
\\&+\theta_3w_z^2v_s+(2\theta_{12}+2\theta_{15})vv_zw_s+2\theta_{14}v_zw_sw_z+(2\theta_5+2\theta_8)vw_s^2+2\theta_7w_zw_s^2)
\end{aligned}
\eqno(3.5)
$$
$$
\begin{aligned}
&+s^2(\theta_5v_z^2v_s+(2\theta_1+2\theta_2)vv_s^2+2\theta_2w_zv_s^2+4\theta_{12}v_sv_zw_s+4\theta_{5}v_sw_s^2)+s^3\frac{4}{3}\theta_1v_s^3]|_{s=\nu}=0,
\end{aligned}
$$
$$
\begin{aligned}
&\xi_3v_z+2\xi_3w_s+\epsilon[(\eta_7+\eta_8)vv_z+\eta_7v_zw_z+(2\eta_4+2\eta_5)vw_s+2\eta_4w_zw_s
\\&+s(2\eta_7v_zv_s+4\eta_4v_sw_s)]+\epsilon^2[(\frac{1}{2}\theta_{12}+\frac{1}{2}\theta_{13}+\theta_{15})v^2v_z+(\theta_{14}+\theta_{15})vv_zw_z
\\&+\frac{1}{2}\theta_{12}v_zw_z^2+(\theta_5+\theta_6+2\theta_8)v^2w_s+(2\theta_7+2\theta_8)vw_zw_s+\theta_5w_z^2w_s
\\&+s(\frac{1}{6}\theta_{16}v_z^3+(2\theta_{12}+2\theta_{15})vv_zv_s+2\theta_{14}v_zw_zv_s+\theta_{17}v_z^2w_s+(4\theta_5+4\theta_8)vv_sw_s
\\&+4\theta_7w_zv_sw_s+2\theta_{16}v_zw_s^2+\frac{4}{3}\theta_9w_s^3)+s^2(2\theta_{12}v_zv_s^2+4\theta_5v_s^2w_s)]|_{s=\nu}=0,
\end{aligned}
\eqno(3.6)
$$
where $\nu=a^2/l^2$ (square of the diameter-length ratio) is a small
parameter for a slender cylinder. We note that in the above
equations the dependence on $R$ and $a/l$ is entirely through
$s=R^2/l^2$ and $\nu=a^2/l^2$, respectively.

Equations (3.3)-(3.6) comprise a new system of complicated nonlinear
PDE's with complicated boundary conditions, which is still very
difficult to solve directly. However, it is characterized by a small
variable $s$ ($0\leq s\leq \nu$) and two small parameters $\epsilon$
and $\nu$, which permit us to use expansion methods to proceed
further.

%%%%%%%%%%%%%%%%%%%%%%%%%%%%%%%%%%%%%%%%%%%%%%%%%%%%%%%%%%%%%%%%%%%%%%%%%%%%%%%%%%%%%%%%%%%%%%%%%%%
\section{Coupled series-asymptotic expansions}

From equations (3.3)-(3.6), we can see that the two unknowns $w$ and
$v$ depend on the variable $z$ and the small variable $s$ and the
small parameters $\epsilon$ and $\nu$; that is
$$
w=w(z,s;\epsilon,\nu),\ \ \ \ v=v(z,s;\epsilon,\nu).\ \ \eqno(4.1)
$$
To go further, we seek series expansions in terms of the small
variable $s$:
$$
v(z,s;\epsilon,\nu)=V_0(z;\epsilon,\nu)+sV_1(z;\epsilon,\nu)+s^2V_2(z;\epsilon,\nu)+\cdots,\
\ \eqno(4.2)
$$
$$
w(z,s;\epsilon,\nu)=W_0(z;\epsilon,\nu)+sW_1(z;\epsilon,\nu)+s^2W_2(z;\epsilon,\nu)+\cdots.\
\ \eqno(4.3)
$$
Substituting (4.2) and (4.3) into (3.3), the left hand becomes a
series in $s$ and all the coefficients of $s^n$ ($n$=$0$, $1$, $2$,
$\cdots$) should vanish. Equating the coefficients of $s^0$ and
$s^1$ to be zero yields that
$$
\begin{aligned}
&4\xi_4W_1+(2\xi_2+2\xi_3)V_{0z}+\xi_1W_{0zz}+\epsilon((4\eta_4+4\eta_5)V_0W_1
\\&+(2\eta_2+2\eta_3+2\eta_7+2\eta_8)V_0V_{0z}+4\eta_4W_1W_{0z}+(2\eta_2+2\eta_7)V_{0z}W_{0Z}
\\&+2\eta_2V_0W_{0zz}+\eta_1W_{0z}W_{0zz})+\epsilon^2H_1(V_0,W_0,W_1)=0,
\end{aligned}
\eqno(4.4)
$$
$$
\begin{aligned}
&16\xi_3W_2+(4\xi_2+4\xi_3)V_{1z}+\xi_1W_{1zz}+\epsilon((24\eta_4+8\eta_5)V_1W_1
\\&+(16\eta_4+16\eta_5)V_0W_2+(4\eta_2+4\eta_3+12\eta_7+4\eta_8)V_1V_{0z}
\\&+(4\eta_2+4\eta_3+4\eta_7+4\eta_8)V_0V_{1z}+16\eta_4W_2W_{0z}+(4\eta_2+4\eta_7)V_{1z}W_{0z}
\\&+12\eta_4W_1W_{1z}+(2\eta_2+6\eta_7)V_{0z}W_{1z}+2\eta_7W_1V_{0zz}+\eta_4V_{0z}V_{0zz}
\\&+4\eta_2V_1W_{0zz}+\eta_1W_{1z}W_{0zz}+2\eta_2V_0W_{1zz}+\eta_1W_{0z}W_{1zz})
\\&+\epsilon^2H_2(V_0,V_1,W_0,W_1,W_2)=0.
\end{aligned}
\eqno(4.5)
$$
Similarly, substituting equations (4.2) and (4.3) into (3.4) and
equating the coefficient of $s^0$ to be zero yields that
$$
\begin{aligned}
&8\xi_1V_1+(2\xi_2+2\xi_3)W_{1z}+\xi_3V_{0zz}+\epsilon((8\eta_1+8\eta_2)V_0V_1
\\&+(6\eta_4-2\eta_5)W_1^2+8\eta_7W_1V_{0z}+(\frac{5}{2}\eta_4+\frac{1}{2}\eta_5)V_{0z}^2
\\&+8\eta_2V_1W_{0z}+(2\eta_2+2\eta_3+2\eta_7+2\eta_8)V_0W_{1z}+(2\eta_2+2\eta_7)W_{0z}W_{1z}
\\&+(\eta_4+\eta_5)V_0V_{0zz}+\eta_4W_{0z}V_{0zz}+2\eta_7W_1W_{0zz}+\eta_4V_{0z}W_{0zz})
\\&+\epsilon^2H_3(V_0,V_1,W_0,W_1)=0.
\end{aligned}
\eqno(4.6)
$$
The expressions of $H_1$, $H_2$ and $H_3$ are very lengthy, which
are omitted for brevity. Substituting (4.2) and (4.3) into the
traction-free boundary conditions (3.5) and (3.6), we obtain
$$
\begin{aligned}
&(\xi_1+\xi_2)V_0+\xi_2W_{0z}+\nu((3\xi_1+\xi_2)V_1+\xi_2W_{1z})+\epsilon((\frac{1}{2}\eta_1+\frac{3}{2}\eta_2)V_0^2
\\&+(\eta_2+\eta_3)V_0W_{0z}+\frac{1}{2}\eta_2W_{0z}^2)+\epsilon^2H_4(V_0,W_0)+\epsilon\nu H_5(V_0,V_1,W_0,W_1)=0,
\end{aligned}
\eqno(4.7)
$$
$$
\begin{aligned}
&2\xi_3W_1+\xi_3V_{0z}+\nu(4\xi_3W_2+\xi_3V_{1z})+\epsilon((2\eta_4+2\eta_5)V_0W_1+(\eta_7+\eta_8)V_0V_{0z}
\\&+2\eta_4W_1W_{0z}+\eta_7V_{0z}W_{0z})+\epsilon^2H_6(V_0,W_0,W_1)+\epsilon\nu H_7(V_0,V_1,W_0,W_1,W_2)=0,
\end{aligned}
\eqno(4.8)
$$
where the lengthy expressions for $H_4$-$H_7$ are omitted, and
interested readers can find the expressions for $H_i$ ($i$ $=$ $1$,
$\cdots$, $7$) in Appendix C. In obtaining the above equations, we
have neglected terms higher than $O(\epsilon\nu,\epsilon^2)$.

Now the governing equations (3.3)-(3.4) are changed into a
one-dimensional system of differential equations (4.4)-(4.8) for the
unknowns $W_0$, $W_1$, $W_2$, $V_0$ and $V_1$. By using a regular
perturbation method, we can express $W_1$ by $W_0$ and $V_0$ from
(4.4), then substitute $W_1$ into (4.6), we can also express $V_1$
by $W_0$ and $V_0$. Similarly, we substitute $W_1$ and $V_1$ into
(4.5) to obtain the expression of $W_2$ in terms of $W_0$ and $V_0$
through the regular perturbation method. The results are given
below:
$$
\begin{aligned}
W_1=&(-\frac{1}{2}-\frac{\xi_2}{2\xi_3})V_{0z}-\frac{\xi_1}{4\xi_3}W_{0zz}+\epsilon(a_1V_0V_{0z}+a_2V_{0z}W_{0z}
\\&+a_3V_{0z}W_{0zz}+a_4W_{0z}W_{0zz})+\epsilon^2(a_5V_0^2V_{0z}+a_6V_0V_{0z}W_{0z}
\\&+a_7V_{0z}W_{0z}^2+a_8V_0^2W_{0zz}+a_9V_0W_{0z}W_{0zz}+a_{10}W{0z}^2W_{0zz}),
\end{aligned}
\eqno(4.9)
$$
$$
\begin{aligned}
V_1=&(\frac{\xi_2}{4\xi_1}+\frac{\xi_2^2}{8\xi_1\xi_3})V_{0zz}+(\frac{1}{16}+\frac{\xi_2}{16\xi_3})W_{0zzz}+\epsilon(a_{11}V_{0z}^2+a{12}V_0V_{0zz}
\\&+a_{13}W_{0z}V_{0zz}+a_{14}V_{0z}W_{0zz}+a_{15}W_{0zz}^2+a_{16}V_0W_{0zzz}+a_{17}W_{0z}W_{0zzz})
\\&+\epsilon^2(a_{18}V_0V_{0z}^2+a_{19}V_{0z}^2W_{0z}+a_{20}V_0^2V_{0zz}+a_{21}V_0W_{0z}V_{0zz}
\\&+a_{22}W_{0z}^2V_{0zz}+a_{23}V_0V_{0z}W_{0zz}+a_{24}V_{0z}W_{0z}W_{0zz}+a_{25}V_0W_{0zz}^2
\\&+a_{26}W_{0z}W_{0zz}^2+a_{27}V_0^2W_{0zzz}+a_{28}V_0W_{0z}W_{0zzz}+a_{29}W_{0z}^2W_{0zzz}),
\end{aligned}
\eqno(4.10)
$$
$$
\begin{aligned}
W_2=&(-\frac{\xi_2}{16\xi_1}+\frac{\xi_1\xi_2}{32\xi_3^2}-\frac{\xi_2^3}{32\xi_1\xi_3^2}+\frac{\xi_1}{32\xi_3}-\frac{3\xi_2^2}{32\xi_1\xi_3})V_{0zzz}+(-\frac{1}{64}+\frac{\xi_1^2}{64\xi_3^2}
\\&-\frac{\xi_2^2}{64\xi_3^2}-\frac{\xi_2}{32\xi_3})W_{0zzzz}+\epsilon(a_{30}V_{0z}V_{0zz}+a_{31}V_{0zz}W_{0zz}+a_{32}V_0V_{0zzz}
\\&+a_{33}W_{0z}V_{0zzz}+a_{34}V_{0z}W_{0zzz}+a_{35}W_{0zz}W_{0zzz}+a_{36}V_0W_{0zzzz}+a_{37}W_{0z}W_{0zzzz})
\\&+\epsilon^2(a_{38}V_{0z}^3+a_{39}V_0V_{0z}V_{0zz}+a_{40}V_{0z}W_{0z}V_{0zz}+a_{41}V_{0z}^2W_{0zz}+a_{42}V_0V_{0zz}W_{0zz}
\\&+a_{43}W_{0z}V_{0zz}W_{0zz}+a_{44}V_{0z}W_{0zz}^2+a_{45}W_{0zz}^3+a_{46}V_0^2V_{0zzz}+a_{47}V_0W_{0z}V_{0zzz}
\\&+a_{48}W_{0z}^2V_{0zzz}+a_{49}V_0V_{0z}W_{0zzz}+a_{50}V_{0z}W_{0z}W_{0zzz}+a_{51}V_0W_{0zz}W_{0zzz}
\\&+a_{52}W_{0z}W_{0zz}W_{0zzz}+a_{53}V_0^2W_{0zzzz}+a_{54}V_0W_{0z}W_{0zzzz}+a_{55}W_{0z}^2W_{0zzzz}),
\end{aligned}
\eqno(4.11)
$$
where $a_i$ $(i=1,2,\cdots,55)$ are constants related to material
constants, whose expressions are omitted for brevity. Substituting
$W_1$, $V_1$ and $W_2$ into (4.7) and (4.8) and omitting the higher
order terms yield the following two equations with only two unknowns
$W_0$ and $V_0$:
$$
\begin{aligned}
&(\xi_1+\xi_2)V_0+\xi_2W_{0z}+\epsilon(b_1V_0^2+b_2V_0W_{0z}+b_3W_{0z}^2)+\epsilon^2(b_4V_0^3+b_5V_0^2W_{0z}
\\&+b_6V_0W_{0z}^2+b_7W_{0z}^3)+\nu(b_8V_{0zz}+b_9W_{0zzz})+\epsilon\nu(b_{10}V_{0z}^2+b_{11}V_0V_{0zz}
\\&+b_{12}W_{0z}V_{0zz}+b_{13}V_{0z}W_{0zz}+b_{14}W_{0zz}^2+b_{15}V_0W_{0zzz}+b_{16}W_{0z}W_{0zzz})=0,
\end{aligned}
\eqno(4.12)
$$
and
$$
\begin{aligned}
&-\xi_2V_{0z}-\frac{\xi_1}{2}W_{0zz}+\epsilon(b_{17}V_0V_{0z}+b_{18}V_{0z}W_{0z}+b_{19}V_0W_{0zz}+b_{20}W_{0z}W_{0zz})
\\&+\epsilon^2(b_{21}V_0^2V_{0z}+b_{22}V_0V_{0z}W_{0z}+b_{23}V_{0z}W_{0z}^2+b_{24}V_0^2W_{0zz}+b_{25}V_0W_{0z}W_{0zz}
\\&+b_{26}W_{0z}^2W_{0zz})+\nu(b_{27}V_{0zzz}+b_{28}W_{0zzzz})+\epsilon\nu(b_{29}V_{0z}V_{0zz}+b_{30}V_{0zz}W_{0zz}
\\&+b_{31}V_0V_{0zzz}+b_{32}W_{0z}V_{0zzz}+b_{33}V_{0z}W_{0zzz}+b_{34}W_{0zz}W_{0zzz}+b_{35}V_0W_{0zzzz}
\\&+b_{36}W_{0z}W_{0zzzz})=0,
\end{aligned}
\eqno(4.13)
$$
where $b_i$ $(i=1,2,\cdots,36)$ are also constants related to
material constants, whose expressions are omitted for brevity.
(Note: Interested readers can contact the corresponding author for
the formulas for $a_i$ and $b_i$.) By using the above two equations,
we can express $V_0$ in terms of $W_0$ and its derivatives with
respect to $z$. From (4.12), we have
$$
\begin{aligned}
V_0=&\frac{-1}{\xi_1+\xi_2}[\xi_2W_{0z}+\epsilon(b_1V_0^2+b_2V_0W_{0z}+b_3W_{0z}^2)+\epsilon^2(b_4V_0^3+b_5V_0^2W_{0z}
\\&+b_6V_0W_{0z}^2+b_7W_{0z}^3)+\nu(b_8V_{0zz}+b_9W_{0zzz})+\epsilon\nu(b_{10}V_{0z}^2+b_{11}V_0V_{0zz}
\\&+b_{12}W_{0z}V_{0zz}+b_{13}V_{0z}W_{0zz}+b_{14}W_{0zz}^2+b_{15}V_0W_{0zzz}+b_{16}W_{0z}W_{0zzz})].
\end{aligned}
\eqno(4.14)
$$
From (4.12) and (4.13), we also have
$$
V_0=\frac{-1}{\xi_1+\xi_2}[\xi_2W_{0z}+\epsilon(b_1V_0^2+b_2V_0W_{0z}+b_3W_{0z}^2)+\nu(b_8V_{0zz}+b_9W_{0zzz})]+\cdots,
\eqno(4.15)
$$
and
$$
V_{0z}=\frac{1}{\xi_2}[-\frac{\xi_1}{2}W_{0zz}+\epsilon(b_{17}V_0V_{0z}+b_{18}V_{0z}W_{0z}+b_{19}V_0W_{0zz}+b_{20}W_{0z}W_{0zz})]+\cdots.\
\ \eqno(4.16)
$$
Substituting $V_0=\frac{-\xi_2}{\xi_1+\xi_2}W_{0z}+\cdots$ into
(4.15) and (4.16), then substituting (4.16) into
$\nu(b_8V_{0zz}+b_9W_{0zzz})$ of (4.14) and (4.15) into the other
part of (4.14), we obtain
$$
\begin{aligned}
V_0=&-\frac{\xi_2}{\xi_1+\xi_2}W_{0z}+\nu\frac{\xi_2-\xi_1}{16(\xi_1+\xi_2)}W_{0zzz}+\epsilon\alpha_1W_{0z}^2+\epsilon^2\alpha_2W_{0z}^3
\\&+\epsilon\nu(\alpha_3W_{0zz}^2+\alpha_4W_{0z}W_{0zzz}),
\end{aligned}
\eqno(4.17)
$$
where
$$
\begin{aligned}
\alpha_1=&-\frac{\xi_1^2\eta_2+\xi_2^2(\eta_1+2\eta_2-2\eta_3)-2\xi_1\xi_2\eta_3}{2(\xi_1+\xi_2)^3}.\\
\alpha_2=&\frac{1}{6(\xi_1+\xi_2)^5}[\xi_1^3(3\eta_2^2+3\eta_2\eta_3-\xi_1\theta_2)+3\xi_1\xi_2^2(2\eta_2^2+4\eta_2\eta_3-4\eta_3^2
\\&+\eta_1(\eta_2+3\eta_3)-3\xi_1\theta_2+3\xi_1\theta_3)+\xi_2^4(\theta_1+6\theta_3-6\theta_4)+\xi_2^3(-3\eta_1^2
\\&-12\eta_1\eta_2-12\eta_2^2+9\eta_1\eta_3+18\eta_2\eta_3-6\eta_3^2+\xi_1\theta_1-6\xi_1\theta_2+12\xi_1\theta_3
\end{aligned}
$$
$$
\begin{aligned}
\\&-9\xi_1\theta_4)-\xi_1^2\xi_2(3\eta_1\eta_2+6\eta_2^2+3\eta_2\eta_3+6\eta_3^2+4\xi_1\theta_2-3\xi_1\theta_3-3\xi_1\theta_4)],\\
\alpha_3=&\frac{1}{64\xi_1(\xi_1+\xi_2)^3}[13\xi_1^4+4\xi_2^3(4\eta_1-3\eta_2)-\xi_1^2\xi_2(23\xi_2+\eta_1+13\eta_2-26\eta_3)
\\&+4\xi_1\xi_2^2(3\eta_1-10\eta_2+4\eta_3)+\xi_1^3(10\xi_2-3\eta_1-7\eta_2+6\eta_3)],\\
\alpha_4=&\frac{1}{16\xi_1(\xi_1-\xi_2)(\xi_1+\xi_2)^4}[3\xi_1^6+2\xi_2^5(-\eta_1+\eta_2)+\xi_1^3\xi_2^2(-51\xi_2+10\eta_2-6\eta_3)
\end{aligned}
$$
$$
\begin{aligned}
&+\xi_1^4\xi_2(-21\xi_2-5\eta_1+\eta_2-4\eta_3)+\xi_1^2\xi_2^3(-30\xi_2+7\eta_1+21\eta_2-4\eta_3)
\\&+\xi_1^5(3\xi_2-\eta_1+5\eta_2+2\eta_3)-\xi_1\xi_2^4(7\eta_1+15\eta_2+4\eta_3)].
\end{aligned}
$$
By substituting (4.16) and (4.17) into (4.13), we obtain
$$
\begin{aligned}
&W_{0zz}-\frac{1}{4}\nu W_{0zzzz}+3\epsilon^2D_2W_{0z}^2W_{0zz}
\\&+\epsilon(2D_1W_{0z}W_{0zz}+\nu(4D_3W_{0zz}W_{0zzz}+2D_3W_{0z}W_{0zzzz}))=0,
\end{aligned}
\eqno(4.18)
$$
where
$$
\begin{aligned}
D_1=&\frac{1}{2(\xi_1+\xi_2)^2(\xi_1^2+\xi_1\xi_2-2\xi_2^2)}[\xi_1^3\eta_1+3\xi_1^2\xi_2(\eta_1-2\eta_2)-\xi_2^3(\eta_1+6\eta_2-6\eta_3)
\\&+3\xi_1\xi_2^2(\eta_1-2\eta_2+2\eta_3)],
\end{aligned}
\eqno(4.19)
$$
$$
\begin{aligned}
D_2=&\frac{1}{6(\xi_1+\xi_2)^4(\xi_1^2+\xi_1\xi_2-2\xi_2^2)}[\xi_1^4(-6\eta_2^2+\xi_1\theta_1)+\xi_1^3\xi_2(24\eta_2\eta_3+5\xi_1\theta_1
\\&-8\xi_1\theta_2)+\xi_2^5(3\theta_1-8\theta_2+18\theta_3-12\theta_4)+2\xi_1^2\xi_2^2(-6\eta_1\eta_2-12\eta_2^2+12\eta_2\eta_3
\\&-12\eta_3^3+5\xi_1\theta_1-16\xi_1\theta_2+6\xi_1\theta_3+6\xi_1\theta_4)+2\xi_1\xi_2^3(12\eta_1\eta_3+24\eta_2\eta_3
\\&-24\eta_3^2+5\xi_1\theta_1-28\xi_1\theta_2+18\xi_1\theta_3+6\xi_1\theta_4)-\xi_2^4(6\eta_1^2+24\eta_2^2+24\eta_1\eta_2
\\&-24\eta_1\eta_3-48\eta_2\eta_3+24\eta_3^2-7\xi_1\theta_1+40\xi_1\theta_2-42\xi_1\theta_3+12\xi_1\theta_4)],
\end{aligned}
\eqno(4.20)
$$
$$
\begin{aligned}
D_3=&\frac{1}{8(\xi_1-\xi_2)(\xi_1+\xi_2)^3}[2\xi_1^4+\xi_1^3(4\xi_2-\eta_1-3\eta_2)+\xi_1\xi_2^2(-4\xi_2-7\eta_1+3\eta_2)
\\&+6\xi_2^3(\eta_2-\eta_3)-2\xi_1^2\xi_2(\xi_2+2\eta_1-3\eta_2-3\eta_3)].
\end{aligned}
\eqno(4.21)
$$
Integrating (4.18) once, we obtain
$$
\epsilon W_{0z}-\frac{1}{4}\epsilon\nu
W_{0zzz}+\epsilon^2D_1W_{0z}^2+\epsilon^3D_2W_{0z}^3+\epsilon^2\nu
(D_3W_{0zz}^2+2D_3W_{0z}W_{0zzz})=A,\ \ \ \eqno(4.22)
$$
where $A$ is the integration constant. It is important to find the
physical meaning of $A$, since to capture the instability phenomena
observed in the experiments, one needs to study the global
bifurcation as the physical parameters vary. For that purpose, we
consider the resultant force $T$ acting on the material
cross-section that is planar and perpendicular to the cylinder axis
in the reference configuration, and the formula is
$$
T=\int_0^{2\pi}\int_0^a \Sigma_{Zz} R dR d\theta.\ \ \eqno(4.23)
$$
By using (4.2), (4.3) and the expressions of $V_0$, $V_1$ and $W_1$
in terms of $W_0$ in (2.5), it is possible to express $\Sigma_{Zz}$
in terms of $W_0$. Then, carrying out the integration in (4.23), we
find that
$$
T=\pi a^2 E [\epsilon W_{0z}-\frac{1}{4}\epsilon\nu
W_{0zzz}+\epsilon^2D_1W_{0z}^2+\epsilon^3D_2W_{0z}^3+\epsilon^2\nu
(D_3W_{0zz}^2+2D_3W_{0z}W_{0zzz})],\ \ \ \eqno(4.24)
$$
where $E=\frac{\xi_1^2+\xi_1\xi_2-2\xi_2^2}{\xi_1+\xi_2}$ is the
Young's modulus. Comparing (4.22) and (4.24), we have
$A=\frac{T}{\pi a^2 E}$. If we retain the original dimensional
variable and let $V=W_{0Z}=\epsilon \widetilde{w}_{0z}$, we have
$$
V+D_1V^2+D_2V^3+a^2(-\frac{1}{4}V_{ZZ}+D_3V_Z^2+2D_3VV_{ZZ})=\gamma,\
\ \eqno(4.25)
$$
where $\gamma=\frac{T}{\pi a^2 E}$ is the dimensionless engineering
stress. Since (4.25) is derived from the three-dimensional field
equations, once its solution is found, the three-dimensional strain
and stress fields can also be found. Also, it contains all the
required terms to yield the leading-order behavior of the original
system. Therefore, we refer (4.25) as the normal form equation of
the system of nonlinear PDE's (3.3) and (3.4) together with the
boundary conditions (3.5) and (3.6) under a given axial resultant.

\noindent\textbf{Remark}: Although the final form of the normal form
equation is one dimensional, it is different from other
one-dimensional phenomenological models in literature (see Coleman
1983, Tong \emph{et al.} 2001, Shaw 2002, Chang \emph{et al.} 2006)
in the sense that it is derived from the three-dimensional field
equations in a mathematically consistent manner and all
three-dimensional quantities can be calculated once its solutions
are found. Thus, we can provide the three-dimensional current
configuration in a two-phase state, which cannot be deduced in the
existing one-dimensional models in literature. This equation
contains a higher-order derivative term which plays the role of
regularization. Trunskinovsky (1982, 1985) pioneered the idea of the
regularization augmentation for solid-solid phase transitions which
involves adding terms (like a strain gradient) into the usual
constitutive stress-strain relation (for the gradient approach, see
also Aifantis $\&$ Serrin 1983 and Triantafyllidis $\&$ Aifantis
1986). Different from these gradient theories, the gradient term
(which represents the influence of the radial deformation and
traction-free boundary conditions) in our equation is derived and
its coefficient is explicitly given.

%%%%%%%%%%%%%%%%%%%%%%%%%%%%%%%%%%%%%%%%%%%%%%%%%%%%%%%%%%%%%%%%%%%%%%%%%%%%%%%%%%%%%%%%%%%%

\section{The Euler-Lagrange equation}

The normal form equation (4.25) is derived in the previous section
based on the equilibrium equations (2.7) and (2.8). Now, we consider
the energy and show that the same equation can be derived through
the variational principle. The calculations are tremendously
complicated in the present method (we use Mathematica for all the
symbolic computations), and if the two derivations of the normal
form equation agree with each other, it also supports the
correctness of our calculations.

We expand the strain energy function up to the fourth-order
nonlinearity (which implies that the stress components are up to the
third-order nonlinearity), and as a result we obtain
$$
\begin{aligned}
\Phi=&\frac{1}{2}(\xi_1\frac{U^2}{R^2}+\xi_3U_Z^2+2\xi_2\frac{U}{R}W_Z+\xi_1W_Z^2+2\xi_2\frac{U}{R}U_R+2\xi_2W_ZU_R+\xi_1U_R^2
\\&+2\xi_3U_ZW_R+\xi_3W_R^2)+\frac{1}{6}(\eta_1\frac{U^3}{R^3}+3\eta_5\frac{U}{R}U_Z^2+3\eta_2\frac{U^2}{R^2}W_Z+3\eta_4U_Z^2W_Z
\\&+3\eta_2\frac{U}{R}W_Z^2+\eta_1W_Z^3+3\eta_2\frac{U^2}{R^2}U_R+3\eta_4U_Z^2U_R+6\eta_3\frac{U}{R}W_ZU_R+3\eta_2W_Z^2U_R
\\&+3\eta_2\frac{U}{R}U_R^2+3\eta_2W_ZU_R^2+\eta_1U_R^3+6\eta_8\frac{U}{R}U_ZW_R+6\eta_7U_ZW_ZW_R
\\&+6\eta_7U_ZU_RW_R+3\eta_5\frac{U}{R}W_R^2+3\eta_4W_ZW_R^2+3\eta_4U_RW_R^2)
\\&+\frac{1}{24}(\theta_1\frac{U^4}{R^4}+6\theta_6\frac{U^2}{R^2}U_Z^2+\theta_9U_Z^4+4\theta_2\frac{U^3}{R^3}W_Z+12\theta_8\frac{U}{R}U_Z^2W_Z\\
&+6\theta_3\frac{U^2}{R^2}W_Z^2+6\theta_5U_Z^2W_Z^2+4\theta_2\frac{U}{R}W_Z^2+\theta_1W_Z^4+6\theta_3W_Z^2U_R^2\\
&+12\theta_8\frac{U}{R}U_Z^2U_R+12\theta_4\frac{U^2}{R^2}W_ZU_R+12\theta_7U_Z^2W_ZU_R+4\theta_2\frac{U^3}{R^3}U_R\\
&+12\theta_4\frac{U}{R}W_Z^2U_R+4\theta_2W_Z^3U_R+6\theta_3\frac{U^2}{R^2}U_R^2+6\theta_5U_Z^2U_R^2+12\theta_4\frac{U}{R}W_ZU_R^2\\
&+4\theta_2\frac{U}{R}U_R^3+4\theta_2W_ZU_R^3+\theta_1U_R^4+12\theta_{13}\frac{U^2}{R^2}U_ZW_R+24\theta_{14}U_ZW_ZU_RW_R\\
&+24\theta_{15}\frac{U}{R}U_ZW_ZW_R+12\theta_{12}U_ZW_Z^2W_R^2+24\theta_{15}\frac{U}{R}U_ZU_RW_R+4\theta_{16}U_Z^3W_R\\
&+12\theta_{12}U_ZU_R^2W_Z+6\theta_6\frac{U^2}{R^2}W_R^2+6\theta_{17}U_Z^2W_R^2+12\theta_8\frac{U}{R}W_ZW_R^2+6\theta_5W_Z^2W_R^2\\
&+12\theta_8\frac{U}{R}U_RW_R^2+12\theta_7W_ZU_RW_R^2+6\theta_5U_R^2W_R^2+4\theta_{16}U_ZW_R^3+\theta_9W_R^4).
\end{aligned}
\eqno(5.1)
$$
The strain energy per unit length is given by
$$
\Psi=\int_0^a\int_0^{2\pi}\Phi R dR d\Theta. \ \ \eqno(5.2)
$$
By using the same manipulations in section 3 and section 4, we can
get the average strain energy over a cross section
$$
\begin{aligned}
\widetilde{\Psi}=&\frac{\Psi}{\pi a^2}
\\=&\epsilon^2[(\xi_1+\xi_2)V_0^2+2\xi_2V_0W_{0z}+\frac{1}{2}\xi_1W_{0z}^2
\\&+\epsilon((\frac{\eta_1}{3}+\eta_2)V_0^3+(\eta_2+\eta_3)V_0^2W_{0z}+\eta_2V_0W_{0z}^2+\frac{\eta_1}{6}W_{0z}^3)
\\&+\epsilon^2((\frac{\theta_1}{12}+\frac{\theta_2}{3}+\frac{\theta_3}{4})V_0^4+(\frac{\theta_2}{3}+\theta_4)V_0^3W_{0z}+(\frac{\theta_3}{2}+\frac{\theta_4}{2})V_0^2W_{0z}^2
\\&+\frac{\theta_2}{3}V_0W_{0z}^3+\frac{\theta_1}{24}W_{0z}^4)+\nu(\frac{\xi_2^2}{4\xi_3}V_{0z}^2+(\frac{\xi_2^2}{2\xi_1}-\frac{\xi_2^2}{4\xi_3}+\frac{\xi_2^3}{4\xi_1\xi_3})V_0V_{0zz}
\\&+(-\frac{\xi_1}{4}+\frac{\xi_2^2}{2\xi_1}-\frac{\xi_1\xi_2}{4\xi_3}+\frac{\xi_2^3}{4\xi_1\xi_3})W_{0z}V_{0zz}+\frac{\xi_1\xi_2}{4\xi_3}V_{0z}W_{0zz}+\frac{\xi_1^2}{16\xi_3}W_{0zz}^2
\\&+(\frac{\xi_1}{8}+\frac{\xi_2}{8}-\frac{\xi_1\xi_2}{8\xi_3}+\frac{\xi_2^2}{8\xi_3})V_0W_{0zzz}+(\frac{\xi_2}{8}-\frac{\xi_1^2}{8\xi_3}+\frac{\xi_2^2}{8\xi_3})W_{0z}W_{0zzz})
\\&+\epsilon\nu(c_1V_0V_{0z}^2+c_2V_{0z}^2W_{0z}+c_3V_0^2V_{0zz}+c_4V_0W_{0z}V_{0zz}+c_5W_{0z}^2V_{0zz}
\\&+c_6V_0V_{0z}W_{0zz}+c_7V_{0z}W_{0z}W_{0zz}+c_8V_0W_{0zz}^2+c_9W_{0z}W_{0zz}^2+c_{10}V_0^2W_{0zzz}
\\&+c_{11}V_0W_{0z}W_{0zzz}+c_{12}W_{0z}^2W_{0zzz})],
\end{aligned}
\eqno(5.3)
$$
where $c_i$ ($i=1,\cdots,12$) are some constants, whose expressions
can be found in Appendix A.

By further using (4.12) and (4.13), we can reduce the above equation
as
$$
\begin{aligned}
\widetilde{\Psi}=&\epsilon^2 E [\frac{1}{2}W_{0z}^2-\frac{1}{8}\nu
W_{0z}W_{0zzz}+\frac{1}{3}D_1\epsilon W_{0z}^3
\\&+\frac{1}{4}D_2\epsilon^2W_{0z}^4+\epsilon\nu(H_1W_{0z}W_{0zz}^2+H_2W_{0z}^2W_{0zzz})]
\\=&E
[\frac{1}{2}V^2+\frac{1}{3}D_1V^3+\frac{1}{4}D_2V^4-\frac{1}{8}a^2
VV_{ZZ}+a^2(H_1VV_Z^2+H_2V^2V_{ZZ})],
\end{aligned}
\eqno(5.4)
$$
where
$$
H_1=\frac{\xi_1^2(\eta_1+2\eta_2)+\xi_2^2(3\eta_1+2\eta_2-2\eta_3)+2\xi_1\xi_2(\eta_1-2\eta_2-2\eta_3)}{8(\xi_1+\xi_2)(\xi_1^2+\xi_1\xi_2-2\xi_2^2)},\
\eqno(5.5)
$$
$$
\begin{aligned}
H_2=&\frac{1}{16(\xi_1+\xi_2)^3(\xi_1^2+\xi_1\xi_2-2\xi_2^2)}[2\xi_1^5+\xi_1^4(8\xi_2-\eta_2)
\\&+\xi_2^4(3\eta_1+14(\eta_2-\eta_3))+2\xi_1^3\xi_2(3\xi_2-\eta_1+\eta_3)
\\&+\xi_1^2\xi_2^2(-8\xi_2-7\eta_1+11\eta_2+2\eta_3)-2\xi_1\xi_2^3(4\xi_2+3\eta_1-6\eta_2+7\eta_3)].
\end{aligned}
\eqno(5.6)
$$
The total potential energy is then given by
$$
\begin{aligned}
\Omega=&\int_{-l}^l\Psi dZ-T\int_{-l}^l V dZ =\pi a^2 (\int_{-l}^l
\widetilde{\Psi} dZ-E\int_{-l}^l\gamma V dZ)
\\=&\pi a^2 E \int_{-l}^l(-\gamma V+\frac{1}{2}V^2+\frac{1}{3}D_1V^3+\frac{1}{4}D_2V^4-\frac{1}{8}a^2 VV_{ZZ}
\\&+a^2(H_1VV_Z^2+H_2V^2V_{ZZ})) dZ.
\end{aligned}
\eqno(5.7)
$$
Further by the variational principle, from the Euler-lagrange
equation we obtain
$$
V+D_1V^2+D_2V^3+a^2[-\frac{1}{4}V_{ZZ}+(2H_2-H_1)V_Z^2+2(2H_2-H_1)VV_{ZZ}]=\gamma.
\ \ \eqno(5.8)
$$
It can be seen from (5.5), (5.6) and (4.21) that $2H_2-H_1=D_3$. So
the equation (5.8) is exactly the equation (4.25).

The expressions of $\tilde{\Psi}$ and $\Omega$ are themselves
important. When there are multiple solutions, the smallest energy
criterion can be used to judge the preferred solution
(configuration).

%%%%%%%%%%%%%%%%%%%%%%%%%%%%%%%%%%%%%%%%%%%%%%%%%%%%%%%%%%%%%%%%%%%%%%%%%%%%%%%%%%%%%%%%%%%%%%

\section{The clamped boundary conditions}

Now, we conduct a detailed analysis on the normal form equation with
clamped end boundary conditions. We suppose that the strain energy
function is non-convex in a homogeneous constant strain state such
that phase transition can take place. This requires
$$
D_1<0,\ \ \ D_2>0,\ \ \ 3D_2<D_1^2<4D_2.\ \ \ \eqno(6.1)
$$
We rewrite equation (4.25) as a first-order system
$$
\begin{aligned}
V_Z&=y\\
y_Z&=\frac{V+D_1V^2+D_2V^3+D_3a^2y^2-\gamma}{a^2(\frac{1}{4}-2D_3V)}.
\end{aligned}
\eqno(6.2)
$$
The critical points of this system are determined by $y=0$ and
$$
V+D_1V^2+D_2V^3=\gamma.\ \ \ \eqno(6.3)
$$

In this paper, we always choose $D_1=-18$, $D_2=100$. We can see
that with these material constants, the critical stress values
$\gamma_1$, $\gamma_2$, $\gamma_m$ (cf. (6.4)) and the corresponding
strain values (cf. Figure 1) are close to the experimental results
(Tse $\&$ Sun 2000 and Li $\&$ Sun 2002). The $\gamma-V$ curve
corresponding to equation (6.3) is shown in Figure 1.
\begin{center}
\includegraphics[width=85mm,height=48mm]{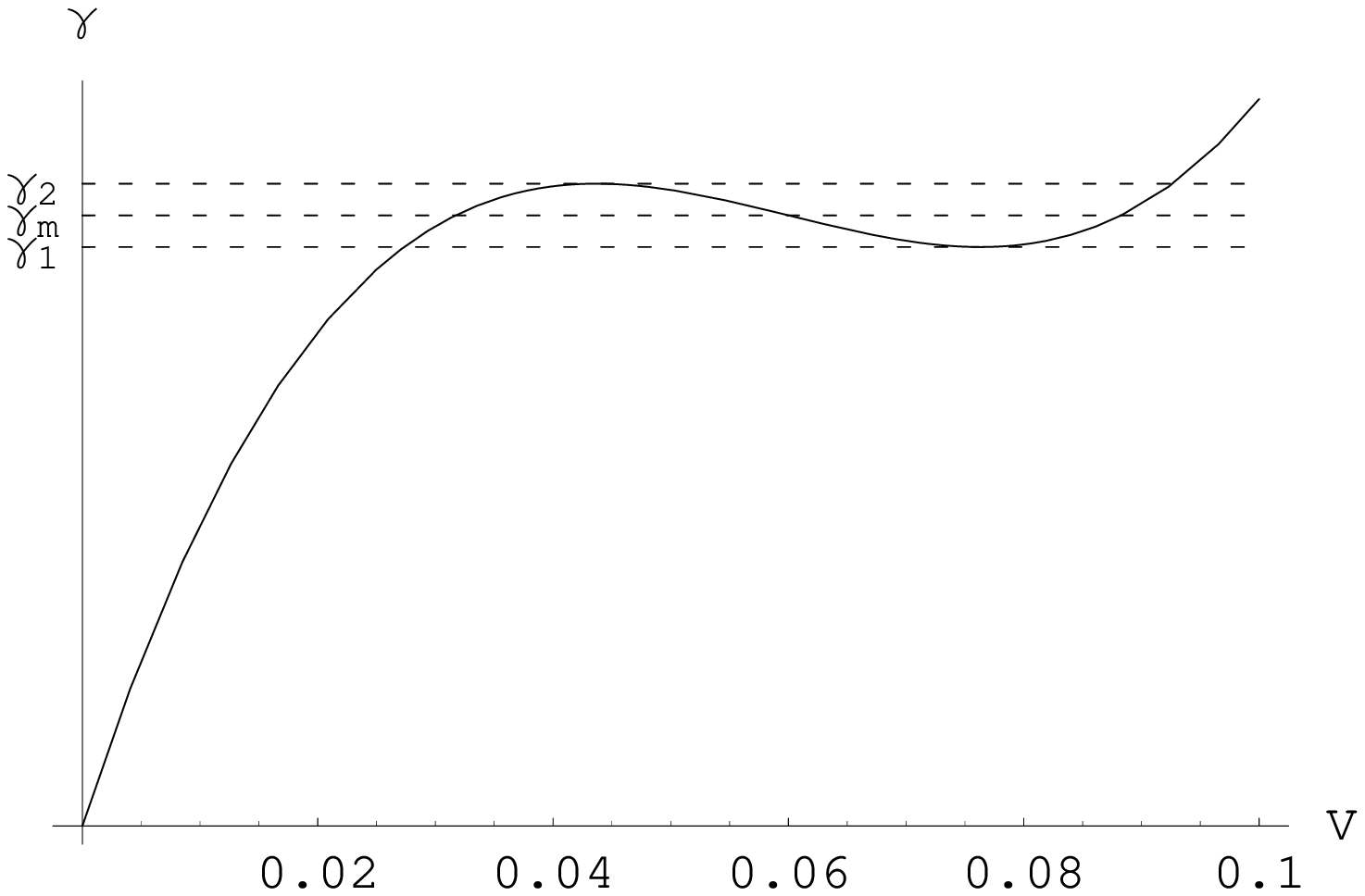}\\
\small\textcolor[rgb]{0.00,0.00,0.00}{Figure 1. The $\gamma-V$ curve
corresponding to (6.3).}
\end{center}
The peak stress value $\gamma_2$, the valley stress value $\gamma_1$
and the Maxwell stress value $\gamma_m$ are
$$
\begin{aligned}
\gamma_1=&\frac{2D_1^3-2(D_1^2-3D_2)^{\frac{3}{2}}-9D_1D_2}{27D_2^2}=0.015929070313677,\
\\
\gamma_2=&\frac{2D_1^3+2(D_1^2-3D_2)^{\frac{3}{2}}-9D_1D_2}{27D_2^2}=0.017670929686323,\
\\ \gamma_m=&\frac{2D_1^3-9D_1D_2}{27D_2^2}=0.016800000000000.
\end{aligned}
\eqno(6.4)
$$

The natural boundary conditions $V_Z=0$ at the two ends have been
used by many authors (e.g., Ericksen 1975; Tong \emph{et al.} 2001;
Cai $\&$ Dai 2006). However, such conditions are difficult to be
realized in practice. In experiments, usually the two ends are
clamped into the rigid bodies. To better describe the realistic
situation, in this paper we will consider the case that the two ends
of the cylinder are clamped into rigid constraints. In this case, we
can propose the conditions according to the fact that there is no
radial displacement for the point at the lateral surface of the
ends, i.e.,
$$
U|_{R=a}=0, \emph{at} \ \ Z=-l, l.\ \ \eqno(6.5)
$$
By using (3.1), (3.2) and (4.2), neglecting the $O(\nu^2)$ terms, we
obtain
$$
V_0+\nu V_1=0, at\ z=-l, l. \ \ \eqno(6.6)
$$
In deriving equation (4.18), $V_0$ and $V_1$ have been expressed in
terms of $W_0$. Using such relations in equation (6.6) and with the
help of equation (4.25), we obtain
$$
F_1 V+F_2 V^2+F_3 V^3=\gamma,\ \ \emph{at} \ \ Z=-l, l.\ \
\eqno(6.7)
$$
where $F_1$, $F_2$ and $F_3$ are material constants. Notice that in
deriving (6.7) we have neglected the terms higher than
$O(\epsilon^2,\nu)$. It's clear that if we choose different values
of $F_1$, $F_2$ and $F_3$, the properties of the boundary condition
(6.7) may also have a lot of differences. Here we will only consider
a simple case. We will choose $F_1$, $F_2$ and $F_3$ such that the
cubic equation (6.7) has only one unique positive root $V_e$ when
$\gamma>0$, and this root is much smaller than the positive roots of
equation (6.3). In this paper, we will choose $F_1=3$, $F_2=-20$ and
$F_3=100$ to get the following graphic result:
\begin{center}
\includegraphics[width=90mm,height=55mm]{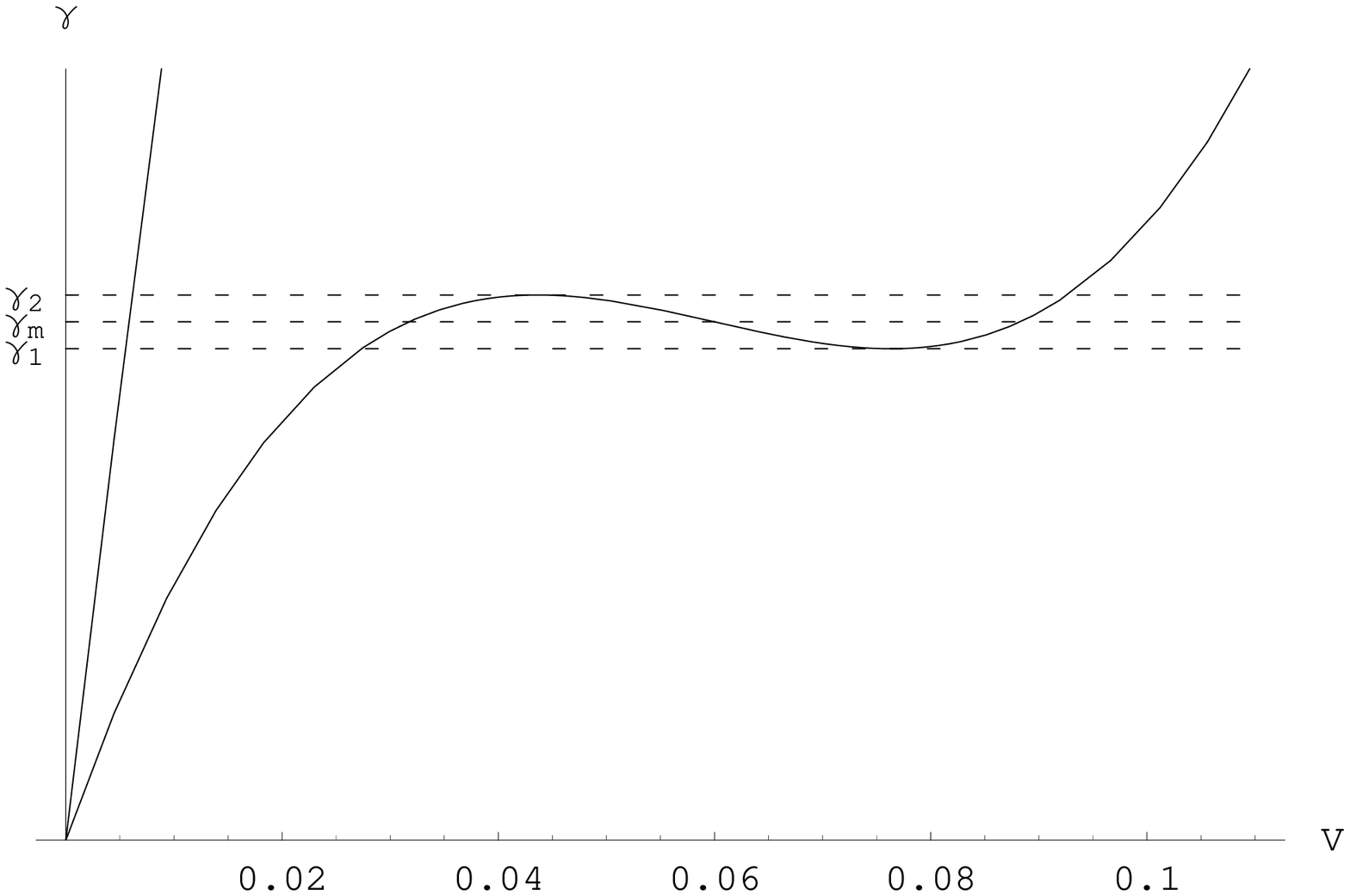}\\
\small\textcolor[rgb]{0.00,0.00,0.00}{Figure 2. Relationship between
the $\gamma-V$ curves corresponding to (6.3) and (6.7).}
\end{center}
Thus the boundary conditions at the two ends are
$$
V=V_e \ \ \  \emph{at}\ \ Z=-l, l.\ \ \eqno(6.8)
$$
We note that $V_e$ depends on the value of the engineering stress
$\gamma$.

\noindent\textbf{Remark}: Although the normal form equation is
one-dimensional, it is derived from a three-dimensional setting, and
as a result we can use the end conditions by considering the
quantities in the radial dimension. If one directly introduces a
one-dimensional model, such an option is not available.

%%%%%%%%%%%%%%%%%%%%%%%%%%%%%%%%%%%%%%%%%%%%%%%%%%%%%%%%%%%%%%%%%%%%%%%%%%%%%%%%%%%%%%%%%%%%%%%%%%%%%%%%%%

\section{Asymptotic solutions}

We now consider the solutions for a given engineering stress
$\gamma$ under the clamped boundary conditions (6.8). In the
following, without loss of generality, we take $l=1$. Then in
equation (4.25), $a$ is equivalent to the diameter-length ratio. Due
to symmetry, we only consider the part $0\leq Z\leq 1$ and boundary
conditions (6.8) can be replaced by
$$
V_Z|_{Z=0}=0,\ \ \ \ V|_{Z=1}=V_e.\ \ \ \eqno(7.1)
$$

\subsection{Force-controlled problem}

We regard the engineering stress $\gamma$ as the bifurcation
parameter. As $\gamma$ varies, there are seven types of phase
planes, which are shown in Figure 3.
\begin{center}
$$
\begin{aligned}
\includegraphics[width=46mm,height=30mm]{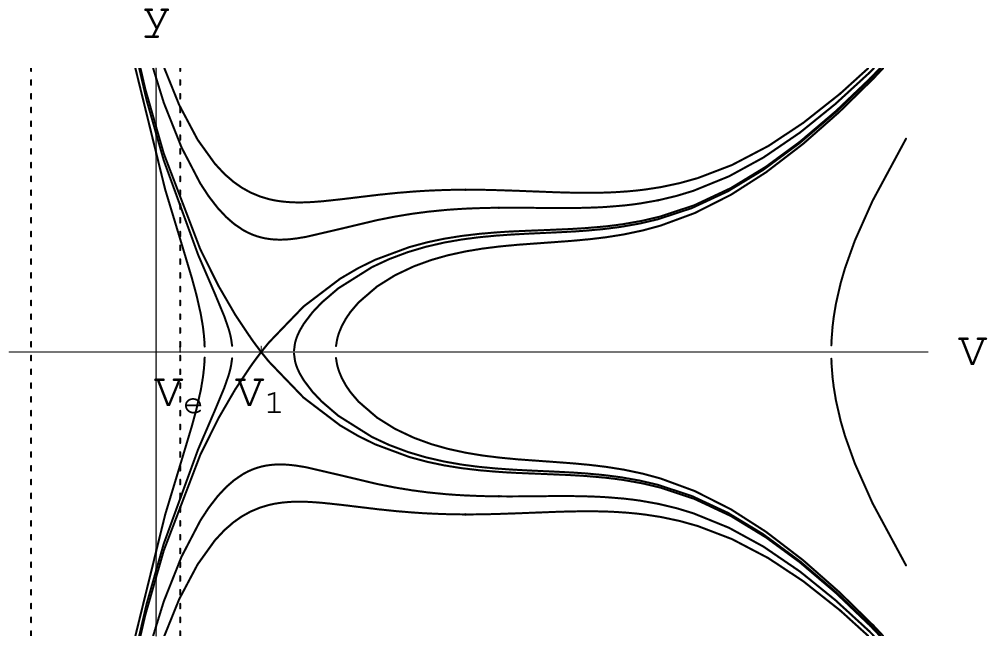}\\
\small\textcolor[rgb]{0.00,0.00,0.00}{(a)}
\end{aligned}\ \ \ \ \ \ \ \ \
\begin{aligned}
\includegraphics[width=46mm,height=30mm]{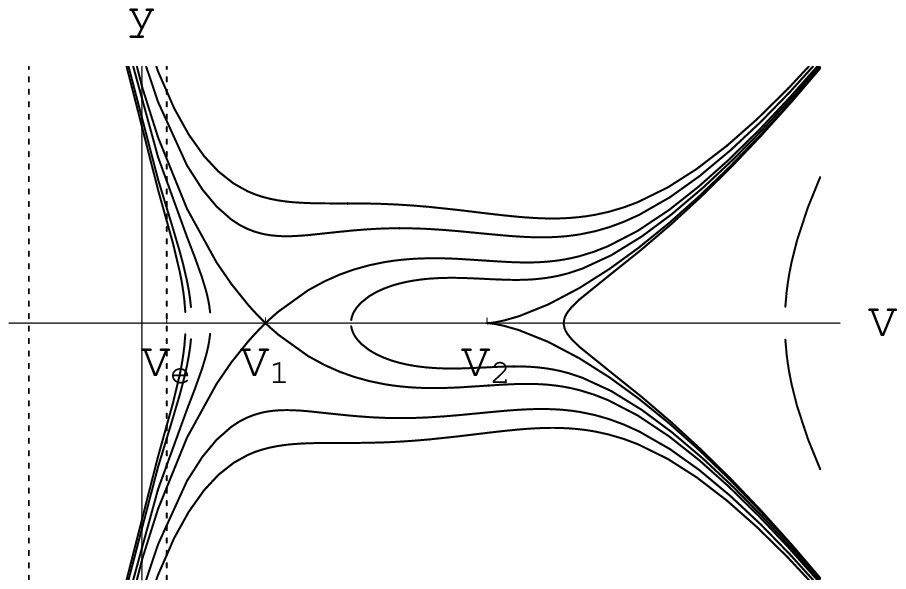}\\
\small\textcolor[rgb]{0.00,0.00,0.00}{(b)}
\end{aligned}
$$
\end{center}
\begin{center}
$$
\begin{aligned}
\includegraphics[width=46mm,height=30mm]{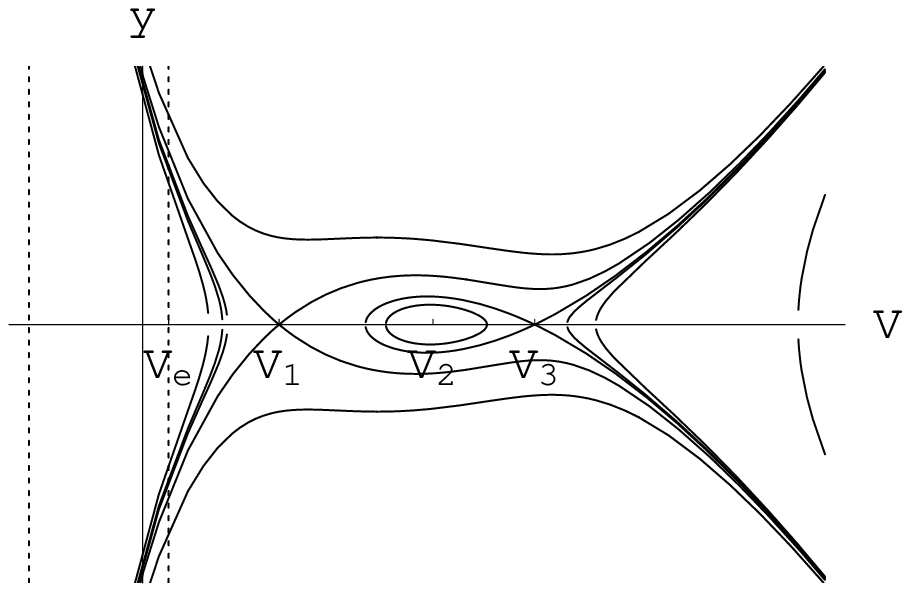}\\
\small\textcolor[rgb]{0.00,0.00,0.00}{(c)}
\end{aligned}\ \ \ \ \ \ \ \ \
\begin{aligned}
\includegraphics[width=46mm,height=30mm]{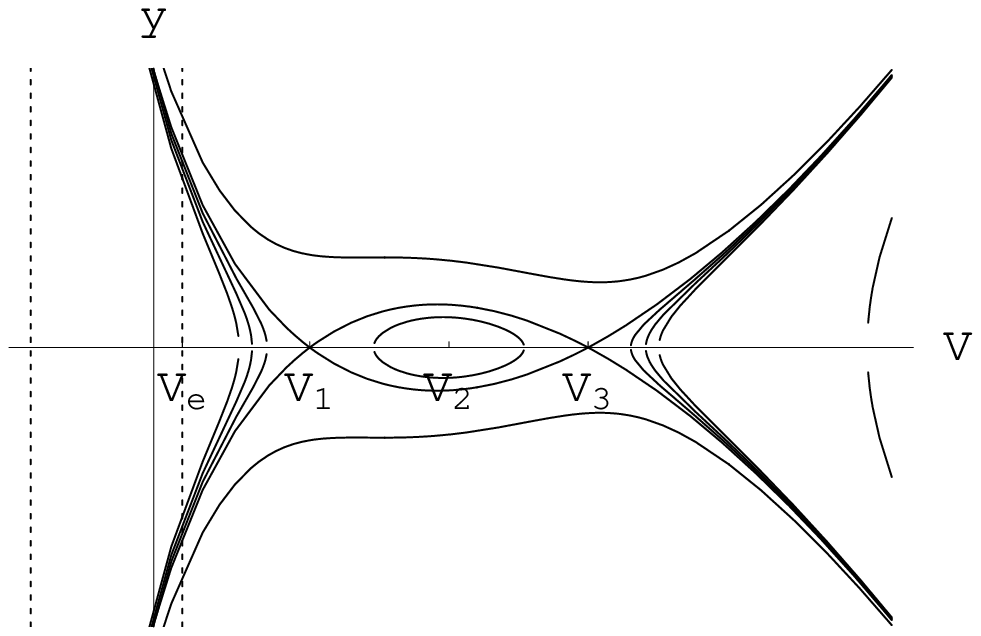}\\
\small\textcolor[rgb]{0.00,0.00,0.00}{(d)}
\end{aligned}
$$
\end{center}
\begin{center}
$$
\begin{aligned}
\includegraphics[width=45.5mm,height=30mm]{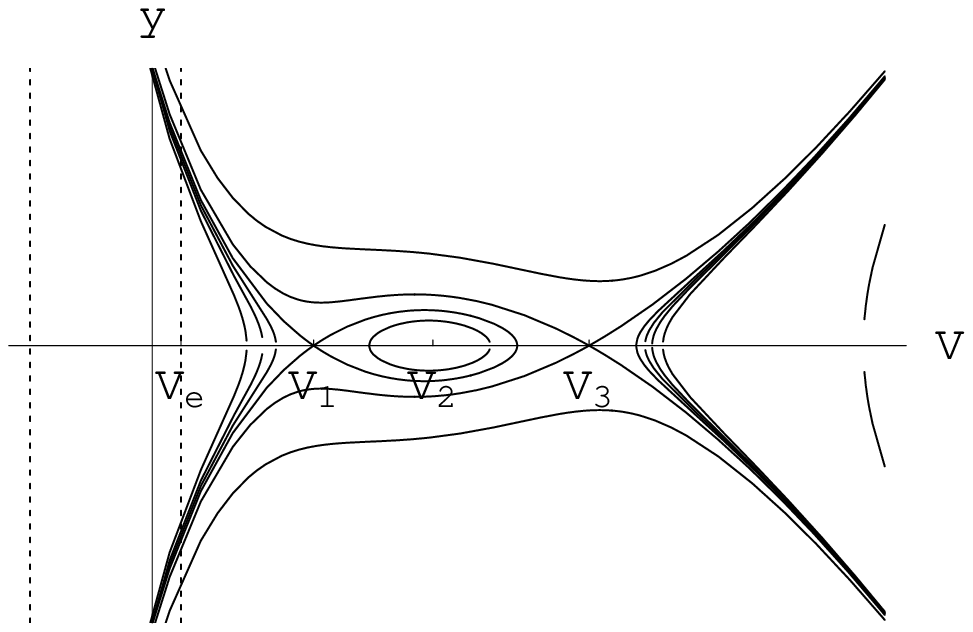}\\
\small\textcolor[rgb]{0.00,0.00,0.00}{(e)}
\end{aligned}\ \ \ \ \ \ \ \ \
\begin{aligned}
\includegraphics[width=45.5mm,height=30mm]{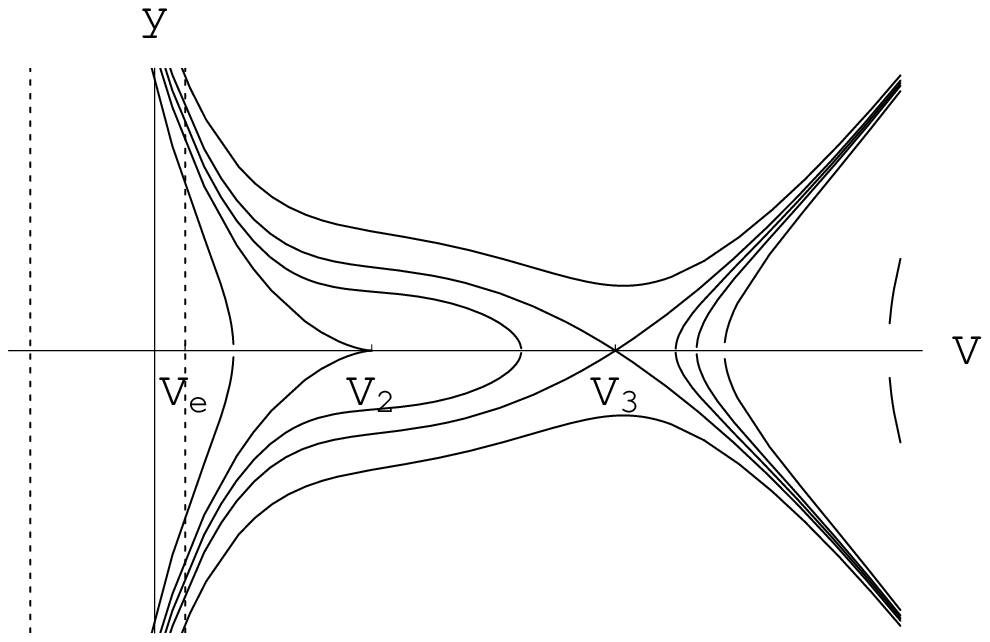}\\
\small\textcolor[rgb]{0.00,0.00,0.00}{(f)}
\end{aligned}\ \ \ \ \ \ \ \ \
\begin{aligned}
\includegraphics[width=45.5mm,height=30mm]{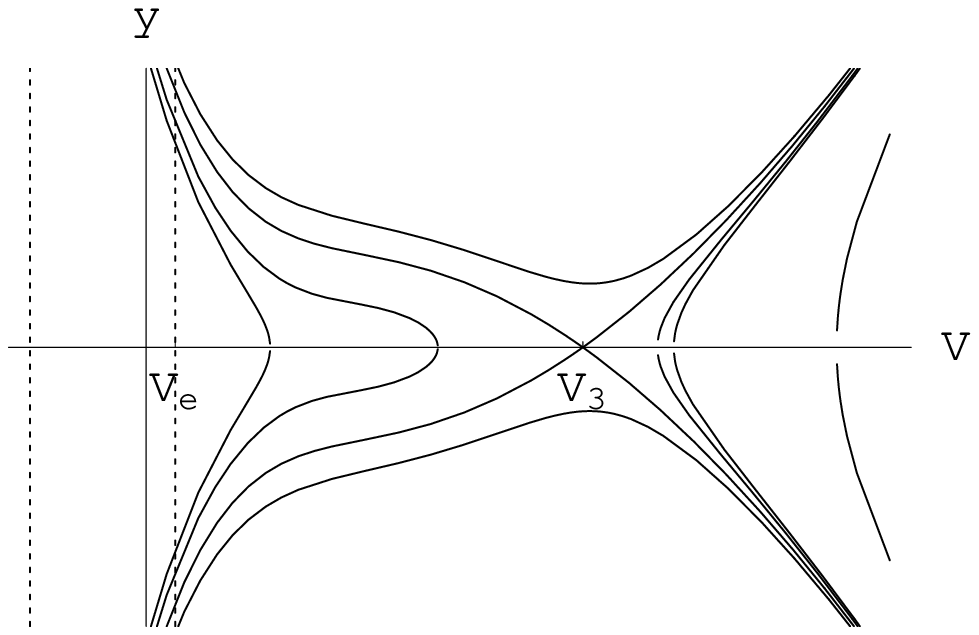}\\
\small\textcolor[rgb]{0.00,0.00,0.00}{(g)}
\end{aligned}
$$
\\
\small\textcolor[rgb]{0.00,0.00,0.00}{Figure 3. Phase planes as the
external stress $\gamma$ varies. (a) $0\leq\gamma\leq\gamma_1$, (b)
$\gamma=\gamma_1$, (c) $\gamma_1<\gamma<\gamma_m$, (d)
$\gamma=\gamma_m$, (e) $\gamma_m<\gamma<\gamma_2$, (f)
$\gamma=\gamma_2$, (g) $\gamma>\gamma_2$.}
\end{center}

Every trajectory in the phase planes of Figure 3 is a solution of
the system (6.2). However, only those trajectories that satisfy
equation (7.1) represent the physical solutions for the present
problem. It can be seen that for a trajectory to be the physical
solution a necessary and sufficient condition is that it contacts
the $V$-axis once (so that $V_Z|_{Z=0}=0$) and contacts the vertical
line $V=V_e$ once (so that $V|_{Z=1}=V_e$) and the Z-interval for
this segment of the trajectory is exactly equal to 1.

To deduce the solution, we integrate (4.25) once to obtain
$$
H-\gamma
V+\frac{1}{2}V^2+\frac{1}{3}D_1V^3+\frac{1}{4}D_2V^4+a^2(-\frac{1}{8}V_Z^2+D_3VV_Z^2)=0,\
\ \eqno(7.2)
$$
where $H$ is the integration constant. If one trajectory contact the
$V$-axis at point $(v_0, 0)$, $v_0$ must be a real root of the
equation
$$
H-\gamma V+\frac{1}{2}V^2+\frac{1}{3}D_1V^3+\frac{1}{4}D_2V^4=0\ \
\eqno(7.3)
$$
By using (7.2) and consider the case $0\leq Z \leq 1$, we can get
$$
\begin{aligned}
\frac{dV}{dZ}&=-\sqrt{\frac{H-\gamma
V+\frac{1}{2}V^2+\frac{1}{3}D_1V^3+\frac{1}{4}D_2V^4}{a^2(\frac{1}{8}-D_3V)}}
\\&=-\sqrt{\frac{-D_2}{4a^2D_3}}\sqrt{\frac{\frac{4H}{D_2}-\frac{4\gamma}{D_2}V+\frac{2}{D_2}V^2+\frac{4D_1}{3D_2}V^3+V^4}{V-\beta}},
\end{aligned}
\ \ \eqno(7.4)
$$
with $\beta=\frac{1}{8D_3}$. Here, for simplicity, we choose
$D_3=-5$ such that $\beta < 0$ for graphic results. Suppose that
$V|_{Z=0}=v_0$, we can obtain
$$
Z=-2a\sqrt{\frac{D_3}{-D_2}}\int_{v_0}^V\sqrt{\frac{\tau-\beta}{\frac{4H_0}{D_2}-\frac{4\gamma}{D_2}\tau+\frac{2}{D_2}\tau^2+\frac{4D_1}{3D_2}\tau^3+\tau^4}}d\tau,
\ \ \ \ V_e\leq V\leq v_0,\ \ \ \eqno(7.5)
$$
where
$$
H_0=-(-\gamma
v_0+\frac{1}{2}v_0^2+\frac{1}{3}D_1v_0^3+\frac{1}{4}D_2v_0^4).
$$
To satisfy the end boundary condition (7.1), we need to choose $v_0$
such that
$$
-2a\sqrt{\frac{D_3}{-D_2}}\int_{v_0}^{V_e}\sqrt{\frac{\tau-\beta}{\frac{4H_0}{D_2}-\frac{4\gamma}{D_2}\tau+\frac{2}{D_2}\tau^2+\frac{4D_1}{3D_2}\tau^3+\tau^4}}d\tau=1.\
\ \ \eqno(7.6)
$$
Based on the phase planes in Figure 3 and equation (7.5) and (7.6),
we can construct all the possible solutions for a given $\gamma$.

Case (a) $0<\gamma\leq \gamma_m$ (cf. Figure 3(a)-3(d))

In this case, from Figure 3(a)-3(d), we can see that only the
trajectories located between the left saddle point $V_1$ and the
vertical line $V=V_e$ (notice that $V_e\ll V_1$, cf. Figure 2) can
be the possible solutions.

For any given $\gamma$, $V_e$ and $V_1$ can be calculated
immediately from the two equations (6.3) and (6.7). By using the
equation (7.6), we can determine the value of $v_0$. Once $v_0$ is
determined, through numerical integration, the corresponding
solution can be obtained from (7.5).

For any $\gamma$ satisfies $0<\gamma\leq \gamma_m$, we found that
there is only one solution (denoted by $S_1$), which is very close
to the saddle point $V_1$.

Case (b) $\gamma_m<\gamma<\gamma_2$ (cf. Figure 3(e))

In this case, for some given $\gamma$, there are multiple solutions.

From figure 3(e), we can see that the trajectories located between
the left saddle point $V_1$ and the vertical line $V=V_e$ can be the
possible solutions. Through some calculations, we found that for any
$\gamma_m<\gamma<\gamma_2$, there exist one solution (also denoted
by $S_1$), which is represented by a trajectory to the left of (also
very close to) the saddle point $V_1$.

We can see that the trajectories located between the homoclinic
orbit and the right saddle point $V_3$ can also be the possible
solutions. Through some calculations, we found that there is a
critical engineering stress $\gamma_p=\gamma_p(a)$ which satisfies
$\gamma_m<\gamma_p(a)<\gamma_2$. When $\gamma_p(a)<\gamma<\gamma_2$,
there exist another two solutions. One solution (denoted by $S_2$)
is represented by a trajectory outside (but very close) to the
homoclinic orbit. The other solution (denoted by $S_3$) is
represented by a trajectory outside the second one and to the left
of (very close to) the right saddle point $V_3$.

Thus the critical engineering stress $\gamma_p(a)$ can be considered
as a bifurcation point. When $\gamma_m<\gamma<\gamma_p(a)$, there is
only one solution. When $\gamma_p(a)<\gamma<\gamma_2$, there are
three solutions. We found that $\gamma_p(a)$ is a monotonically
increasing function of the diameter-length ratio $a$, e.g.,
$\gamma_p(0.03)=0.0168005239556$, $\gamma_p(0.06)=0.0168907218423$.

Case (c) $\gamma=\gamma_2$ (cf. Figure 3(f))

In this case, there are three solutions. The first solution (also
denoted by $S_1$) is represented by a trajectory to the left of
(also very close to) the cusp point $V_2$. The second solution (also
denoted by $S_2$) is represented by a trajectory to the right of
(also very close to) the cusp point $V_2$. The third solution (also
denoted by $S_3$) is represented by a trajectory to the left of
(also very close to) the saddle point $V_3$.

Case (d) $\gamma>\gamma_2$ (cf. Figure 3(g))

In this case, there exist another critical engineering stress
$\gamma_q=\gamma_q(a)$ which is a little bit larger than $\gamma_2$.

When $\gamma_2<\gamma<\gamma_q(a)$, there are three solutions. The
first two solutions (also denoted by $S_1$ and $S_2$) are
represented by two trajectories (very close to each other) located
in the middle part of the interval $(V_e,V_3)$. The third solution
(also denoted by $S_3$) is represented by a trajectory to the left
of (very close to) the saddle point $V_3$. When
$\gamma>\gamma_q(a)$, there is only one solution (also denoted by
$S_3$), which is represented by a trajectory to the left of (very
close to) the saddle point $V_3$.

Thus the critical engineering stress $\gamma_q(a)$ can be considered
as another bifurcation point. We found that $\gamma_q(a)$ is also a
monotonically increasing function of the diameter-length ratio $a$,
e.g., $\gamma_q(0.03)=0.017677609814$,
$\gamma_q(0.06)=0.017735345200$.

For the radius $a=0.03$, the axial strain distribution $V(Z)$
corresponding to the above four cases are plotted in Figure 4.
\begin{center}
$$
\begin{aligned}
\includegraphics[width=55mm,height=41mm]{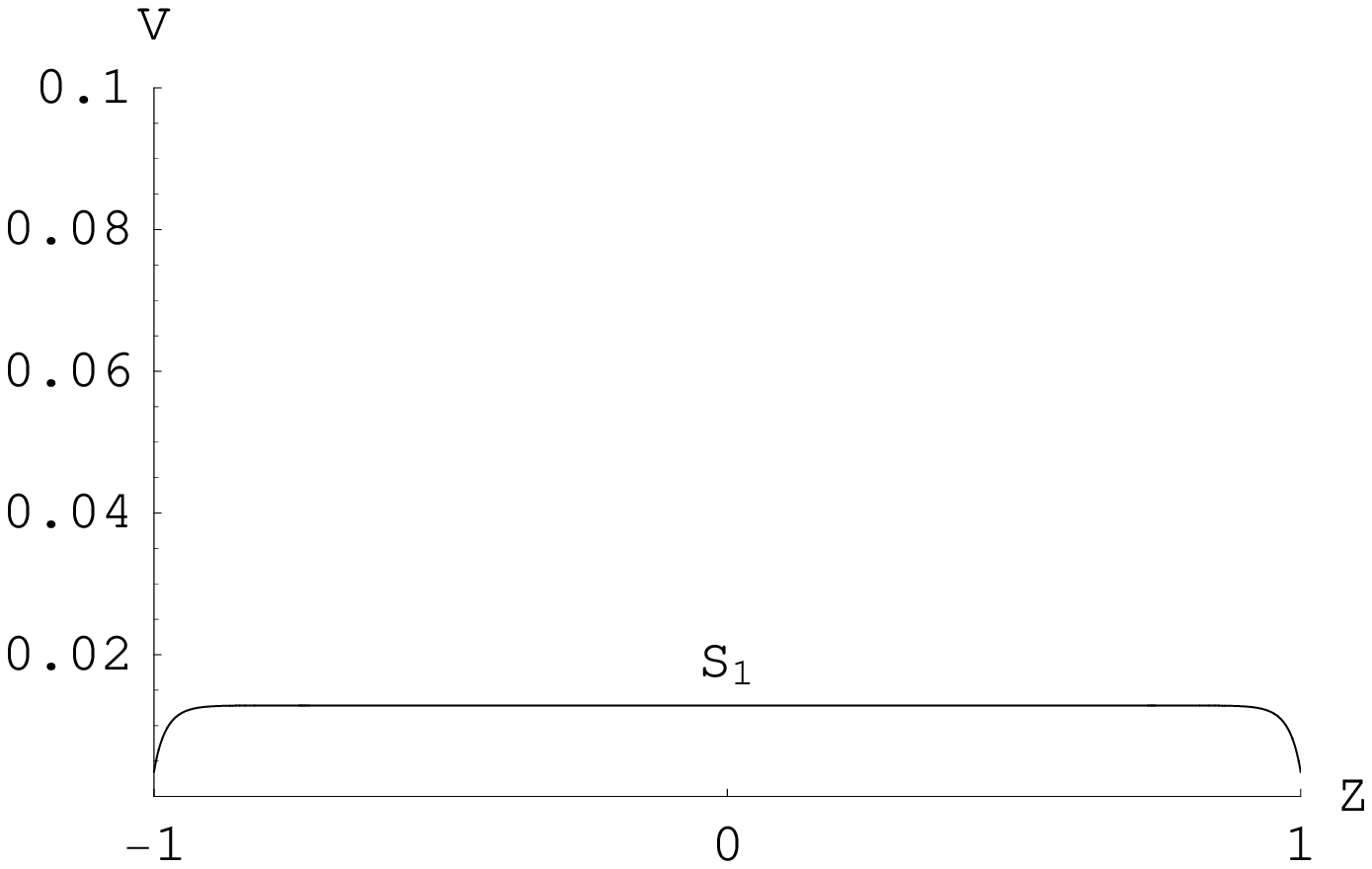}\\
\small\textcolor[rgb]{0.00,0.00,0.00}{(a)}
\end{aligned}\ \ \ \ \ \ \ \ \
\begin{aligned}
\includegraphics[width=55mm,height=41mm]{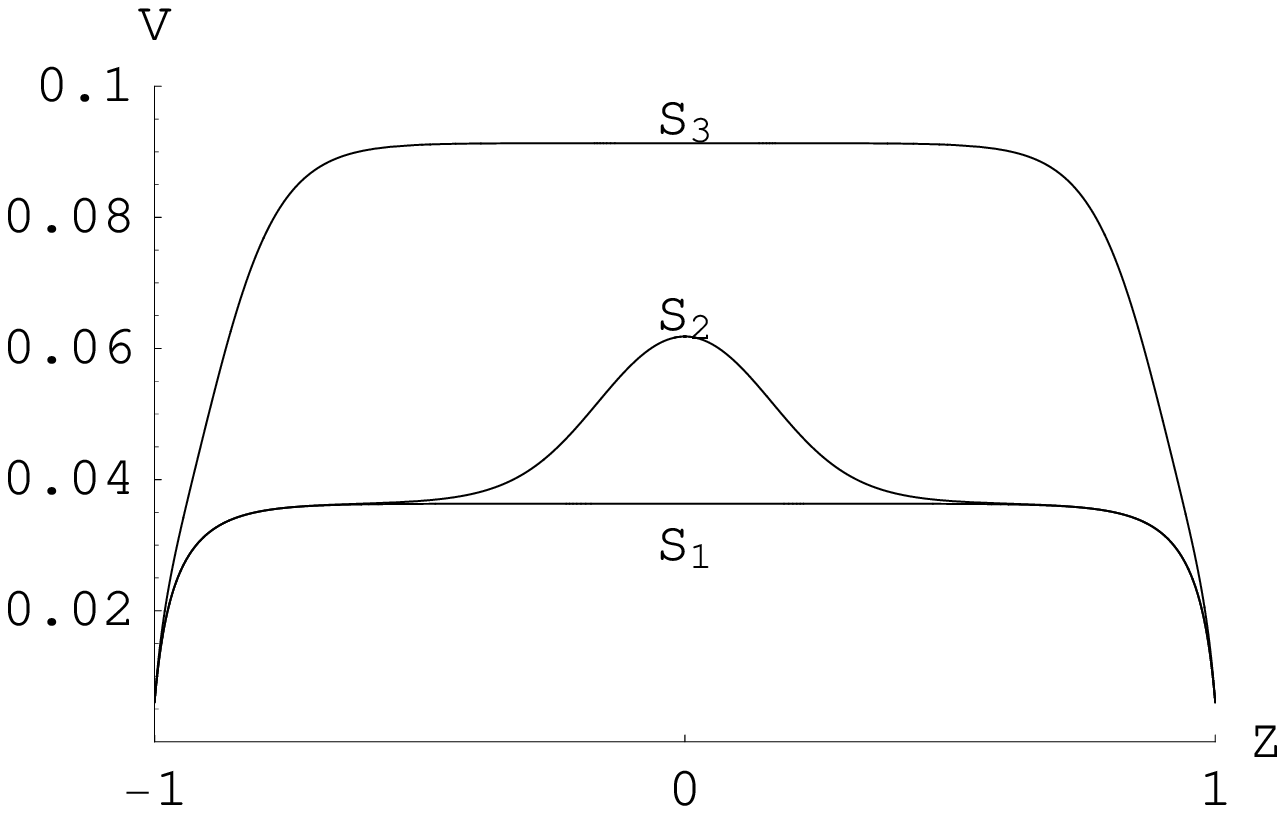}\\
\small\textcolor[rgb]{0.00,0.00,0.00}{(b)}
\end{aligned}
$$
\end{center}
\begin{center}
$$
\begin{aligned}
\includegraphics[width=55mm,height=41mm]{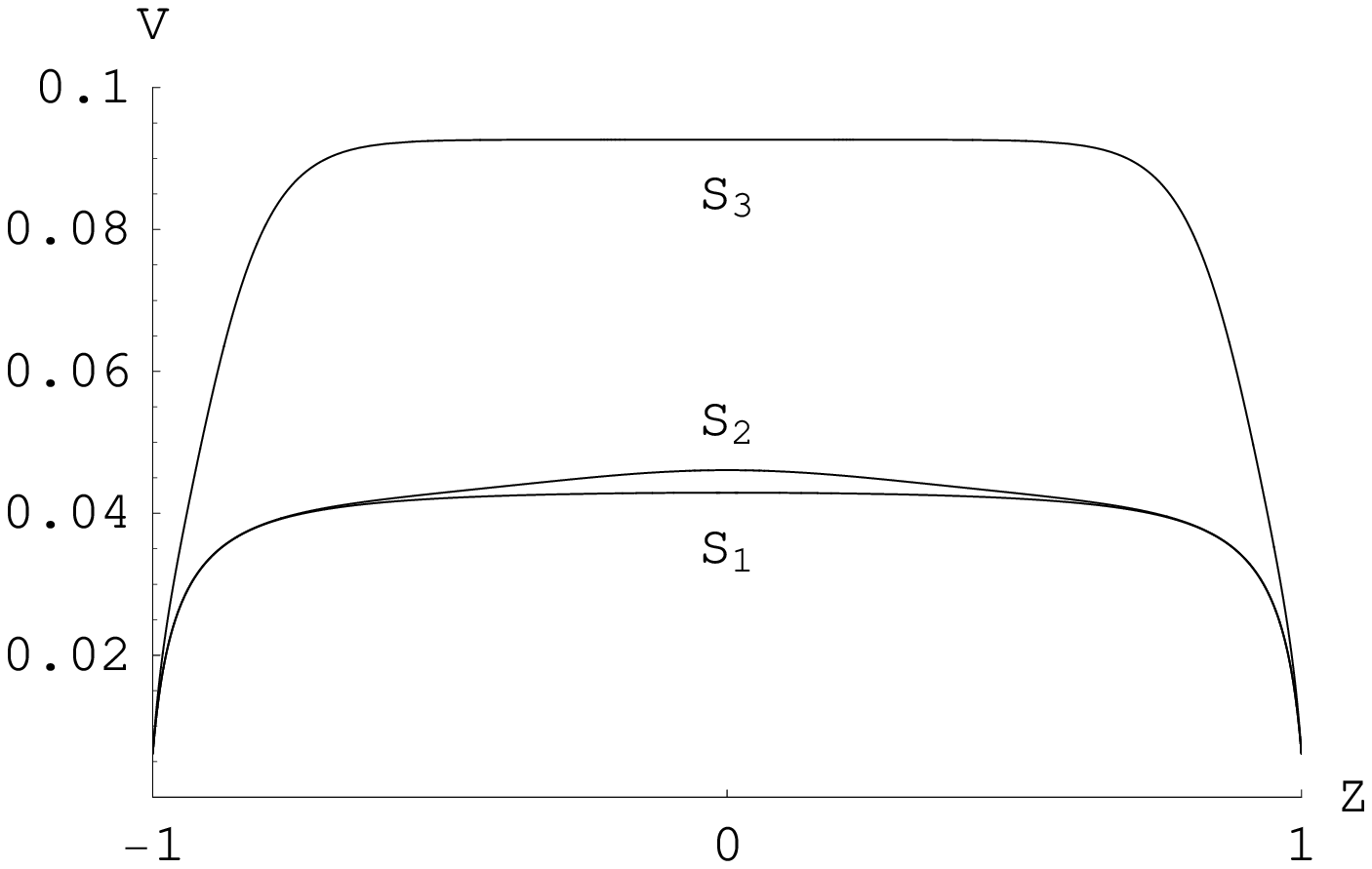}\\
\small\textcolor[rgb]{0.00,0.00,0.00}{(c)}
\end{aligned}\ \ \ \ \ \ \ \ \
\begin{aligned}
\includegraphics[width=55mm,height=41mm]{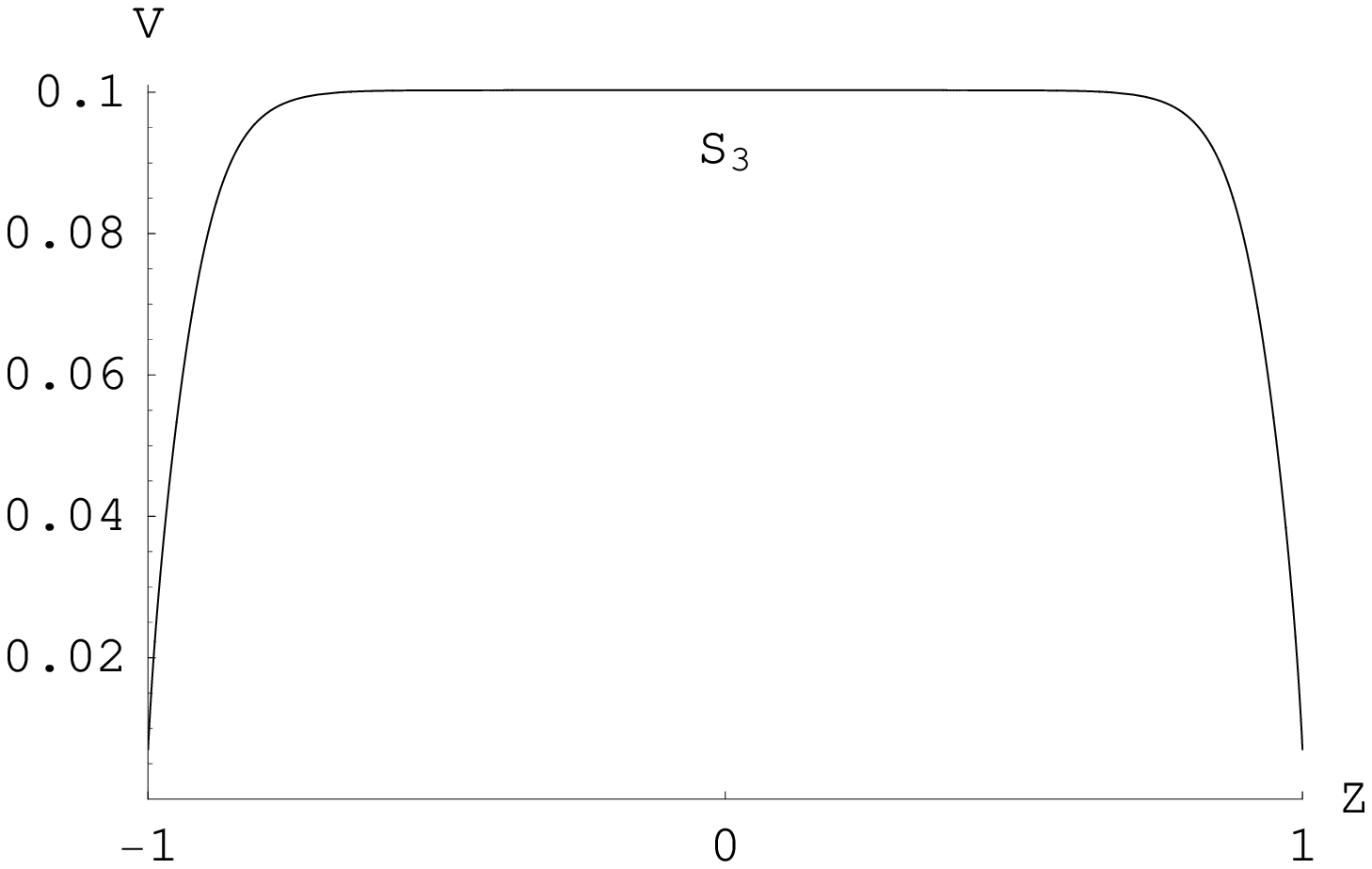}\\
\small\textcolor[rgb]{0.00,0.00,0.00}{(d)}
\end{aligned}
$$
\\
\small\textcolor[rgb]{0.00,0.00,0.00}{Figure 4. The axial strain
distribution curves for four different $\gamma$ values ($a=0.03$):
(a) $0<\gamma=0.010080<\gamma_p$, (b)
$\gamma_p<\gamma=0.017366<\gamma_2$, (c)$\gamma=\gamma_2$, \\(d)
$\gamma=0.020114>\gamma_q$.}
\end{center}

From the above results, it can be seen that there are multiple
solutions for $\gamma_p(a)<\gamma<\gamma_q(a)$. We suppose that the
preferred solution should be the one which has the smallest total
potential energy value. To determine which solution is the most
preferred one, we consider the total potential energy values of all
the possible solutions. From (5.7) and due to symmetric, we can get
$$
\begin{aligned}
\Omega =2\pi a^2 E \int_0^1&(-\gamma
V+\frac{1}{2}V^2+\frac{1}{3}D_1V^3+\frac{1}{4}D_2V^4-\frac{1}{8}a^2
VV_{ZZ}\\&+a^2(H_1VV_Z^2+H_2V^2V_{ZZ})) dZ,
\end{aligned}
\eqno(7.7)
$$
Without loss of generality, we choose $H_1=15$ and $H_2=5$ for
graphic results. For the radius $a=0.03$, we plot the total
potential energies for all the solutions in Figure 5.

\begin{center}
\includegraphics[width=85mm,height=55mm]{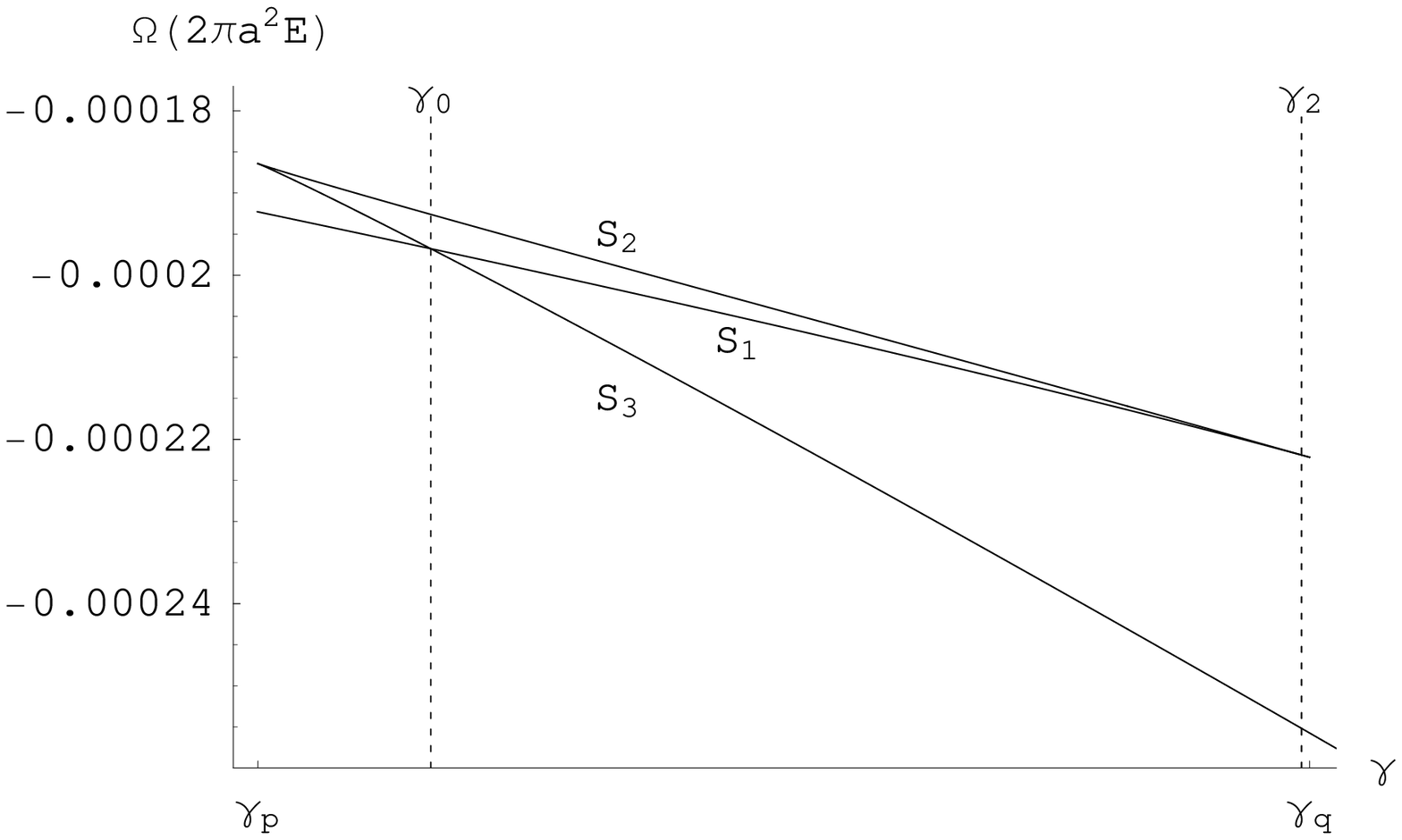}\\
\small\textcolor[rgb]{0.00,0.00,0.00}{Figure 5. Relationships
between $\gamma$ and the potential energies ($a=0.03$).}
\end{center}

From figure 5, it can be seen that there is a critical stress value
$\gamma_0=\gamma_0(a)$ (e.g., $\gamma_0(0.03)=0.01694471757$). For
$\gamma_p<\gamma<\gamma_0$, the first solution $S_1$ is the
preferred solution, and for $\gamma_0<\gamma<\gamma_q$, the third
solution $S_3$ is the preferred solution. Thus, in a loading
process, after $\gamma>\gamma_0$, the configuration of the cylinder
may jump from solution $S_1$ towards solution $S_3$, which is
corresponding to the phase transition process.

\noindent\textbf{Remark}: The existence of $\gamma_0(a)$
($>\gamma_p(a)>\gamma_m$) is significant and it implies that under
clamped end conditions the phase transition can only happen at a
stress value larger than the Maxwell stress.

\subsection{Displacement-controlled problem}

We now consider the case that the total elongation
$$
\int_{-l}^{l}VdZ=2\int_0^{l}VdZ=2\Delta\ \ \ \ \eqno(7.8)
$$
is given. Since we have taken $l$ to be $1$, $\Delta$ is actually
the engineering strain. The governing equation is still equation
(4.25), but now $\gamma$ is an unknown parameter.

In the previous section, we have obtained all the solutions for a
given $\gamma$. If for a given $\Delta$ we can find the
corresponding value $\gamma$, then we can obtain the solutions for a
displacement-controlled problem.

We first plot the $\gamma-\Delta$ curves corresponding to the
solutions we obtained in the previous section for four different
values of $a$ in Figure 6.

\begin{center}
$$
\begin{aligned}
\includegraphics[width=60mm,height=38mm]{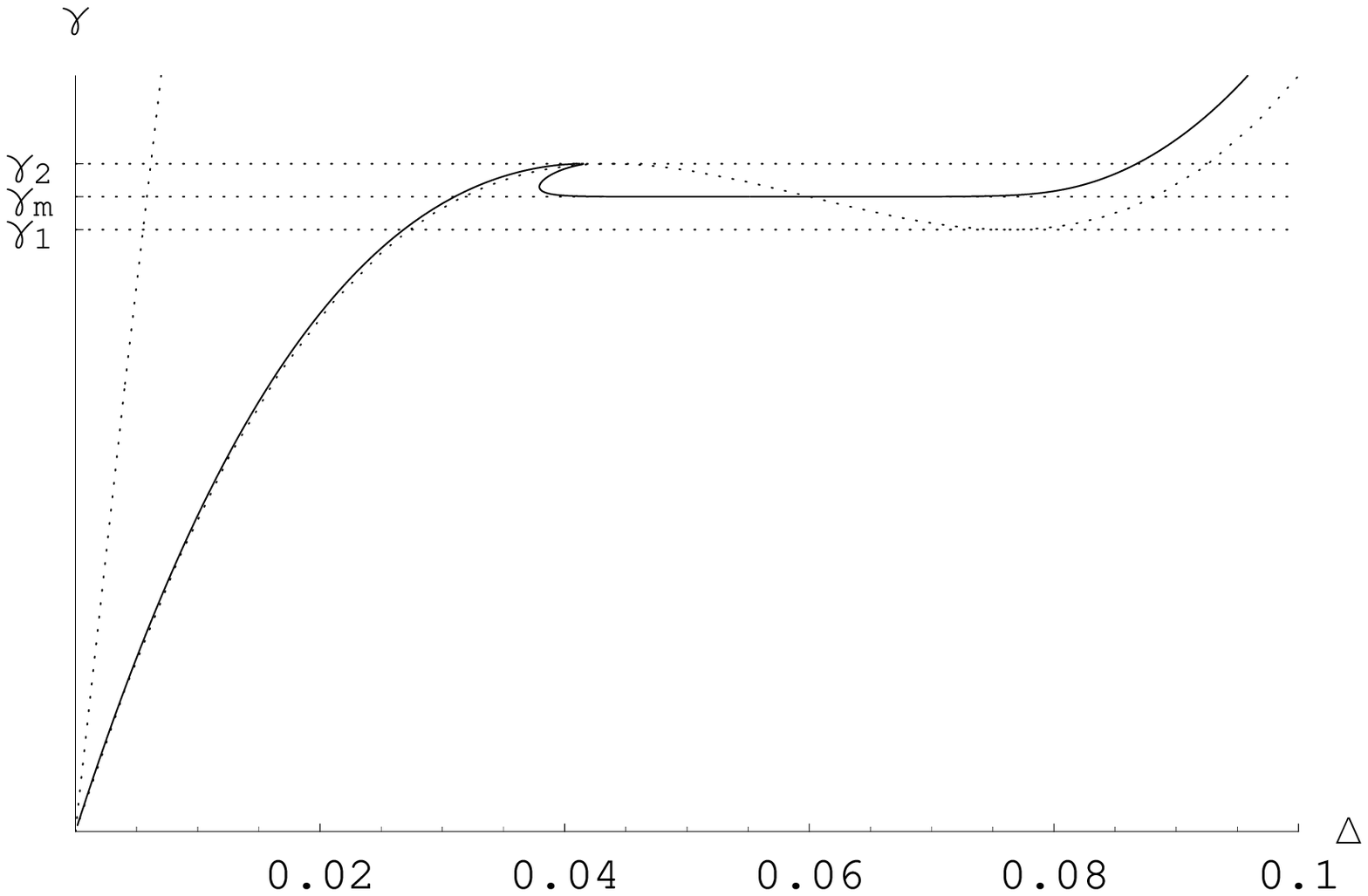}\\
\small\textcolor[rgb]{0.00,0.00,0.00}{(a)}
\end{aligned}\ \ \ \ \ \ \ \ \
\begin{aligned}
\includegraphics[width=60mm,height=38mm]{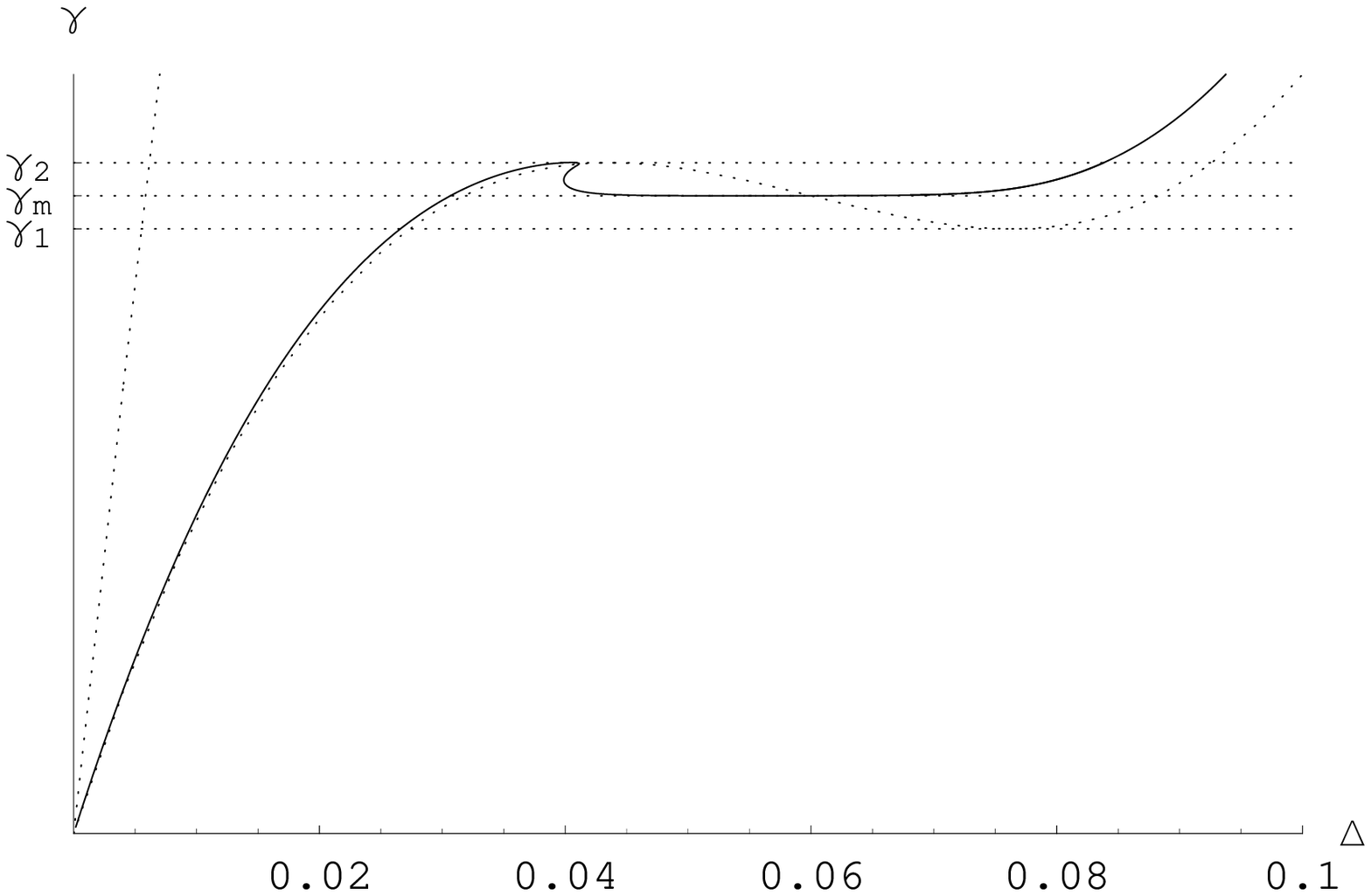}\\
\small\textcolor[rgb]{0.00,0.00,0.00}{(b)}
\end{aligned}
$$
\end{center}
\begin{center}
$$
\begin{aligned}
\includegraphics[width=60mm,height=38mm]{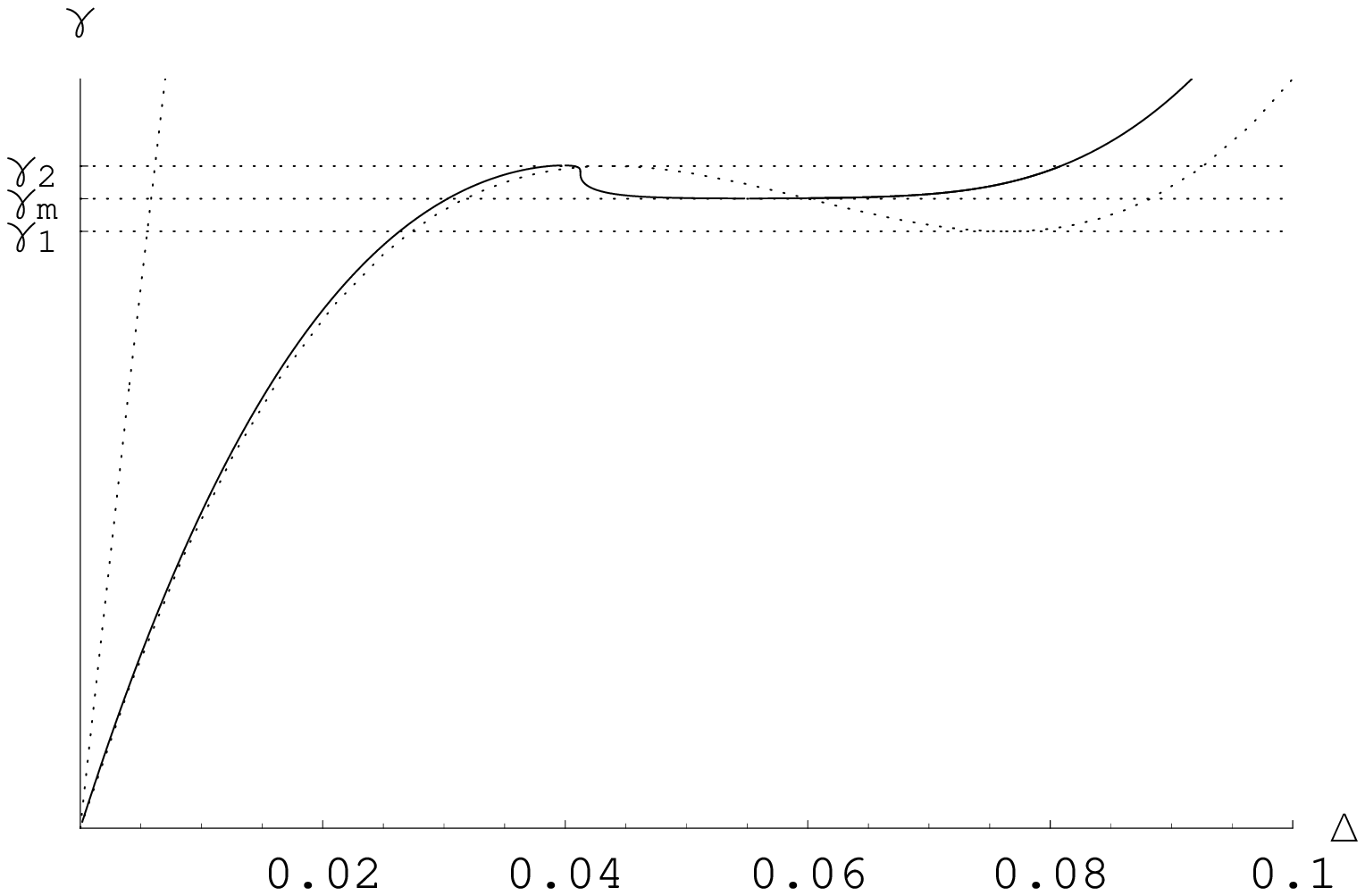}\\
\small\textcolor[rgb]{0.00,0.00,0.00}{(c)}
\end{aligned}\ \ \ \ \ \ \ \ \
\begin{aligned}
\includegraphics[width=60mm,height=38mm]{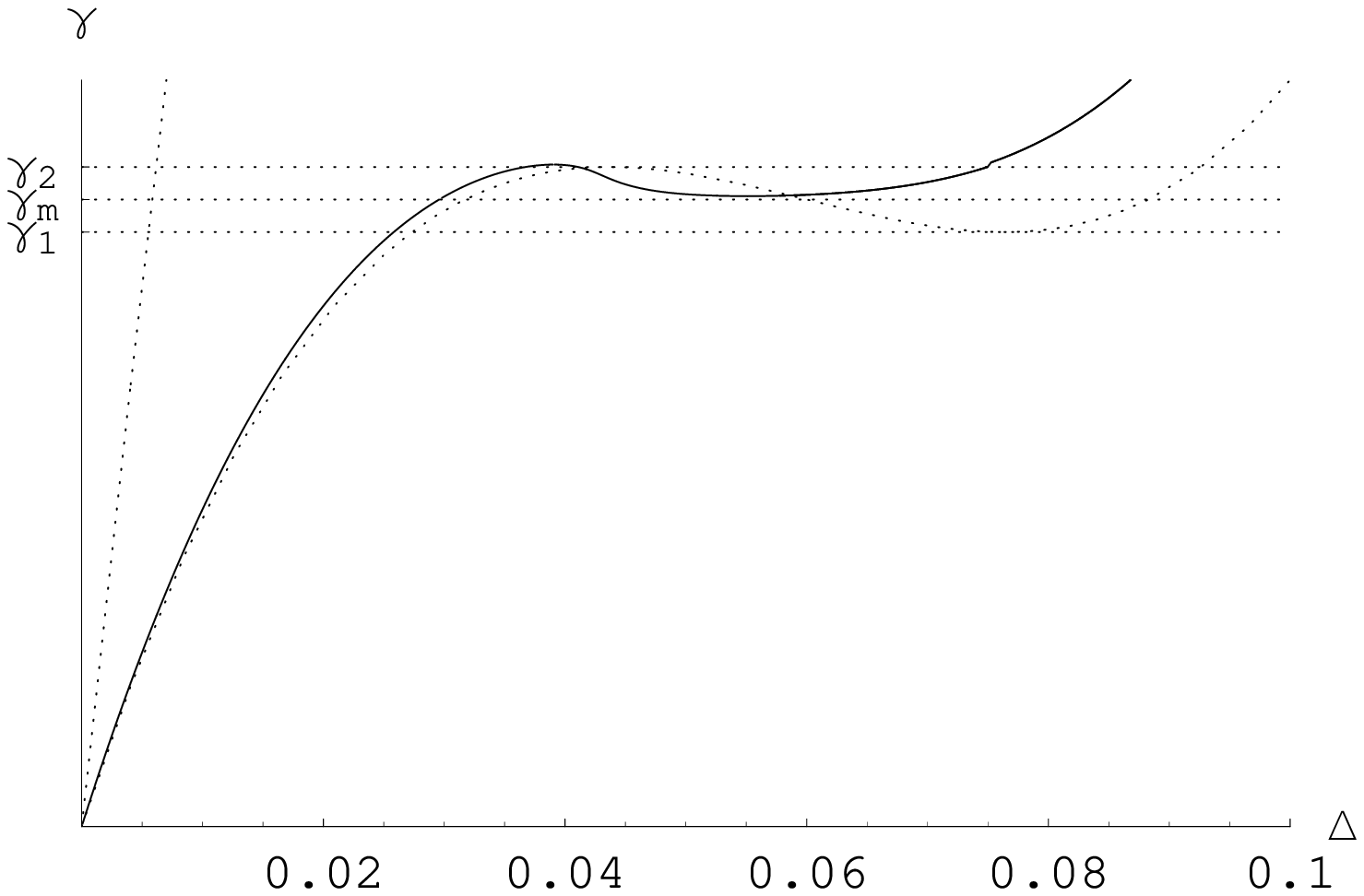}\\
\small\textcolor[rgb]{0.00,0.00,0.00}{(d)}
\end{aligned}
$$
\\
\small\textcolor[rgb]{0.00,0.00,0.00}{Figure 6. The engineering
$\gamma-\Delta$ curves for different values of $a$: (a) $a=0.02$;
(b) $a=0.03$; (c) $a=0.04$; (d) $a=0.06$. The dashed lines are the
curves shown in Figure 2. }
\end{center}

We can see that the $\gamma-\Delta$ curves shown in Figure 6 have
some important features.

First, these curves obtained from the asymptotic solutions capture
the main characteristics of the engineering stress-strain curves
measured in a number of experiments (Shaw $\&$ Kyriakides 1995, Sun
\emph{et al.} 2000, Tse $\&$ Sun 2000, Favier \emph{et al.} 2001 and
Li $\&$ Sun 2002), i.e., there is a stress peak, which is followed
by a sharp stress drop, and then there is a stress plateau (cf.
Figure 6(c)).

Second, it can be seen that there is a snap-back when the
diameter-length ratio $a$ is small (cf. Figure 6(a) and 6(b)). As
the diameter-length ratio become larger, this snap-back phenomenon
disappear (cf. Figure 6(c) and 6(d)).

Third, as the diameter-length ratio $a$ increases, the whole
$\gamma-\Delta$ curve moves towards left, especially in the
high-strain region. It is clear that this feature is due to the
boundary effect.

With these curves, we are ready to convert the solutions of a
force-controlled problem into those of a displacement-controlled
problem. As we pointed out that if the diameter-length ratio $a$ is
large enough, there is no snap-back on the corresponding
$\gamma-\Delta$ curve (cf. Figure 6(c) and 6(d)). Thus, in this
case, for a given displacement $\Delta$, we can only get one
corresponding stress value $\gamma$ from the $\gamma-\Delta$ curve.
For this $\gamma$, there is a unique phase plane. There are maybe
more than one trajectories in this phase plane which satisfy the end
boundary conditions, however only one of them can give the required
displacement value, which then represents the unique solution for
the given $\Delta$.

As we pointed out before that the normal form equation (4.25) is
derived from the three-dimensional field equations. With the
solutions of the normal form equation, we can recover the
three-dimensional strain fields. Here, we take $a=0.04$ as an
example. Corresponding to the four points $A-D$ shown in Figure 7,

\begin{center}
\includegraphics[width=75mm,height=45mm]{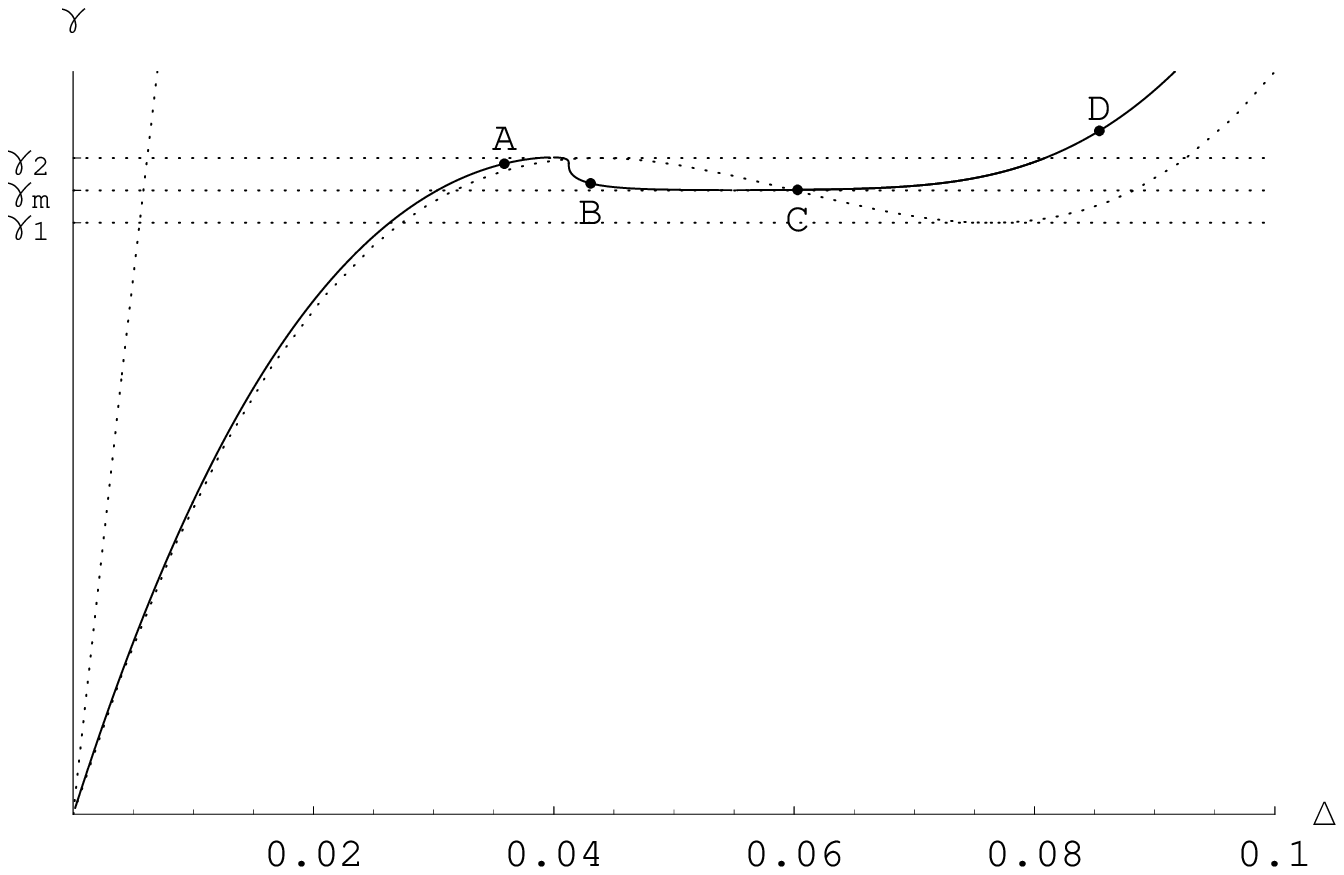}\\
\small\textcolor[rgb]{0.00,0.00,0.00}{Figure 7. The engineering
$\gamma-\Delta$ curve for $a=0.04$.}
\end{center}
we draw the shapes of the cylinder in Figure 8 (the radial
deformation has been enlarged for clearness).

\begin{center}
\includegraphics[width=75mm,height=45mm]{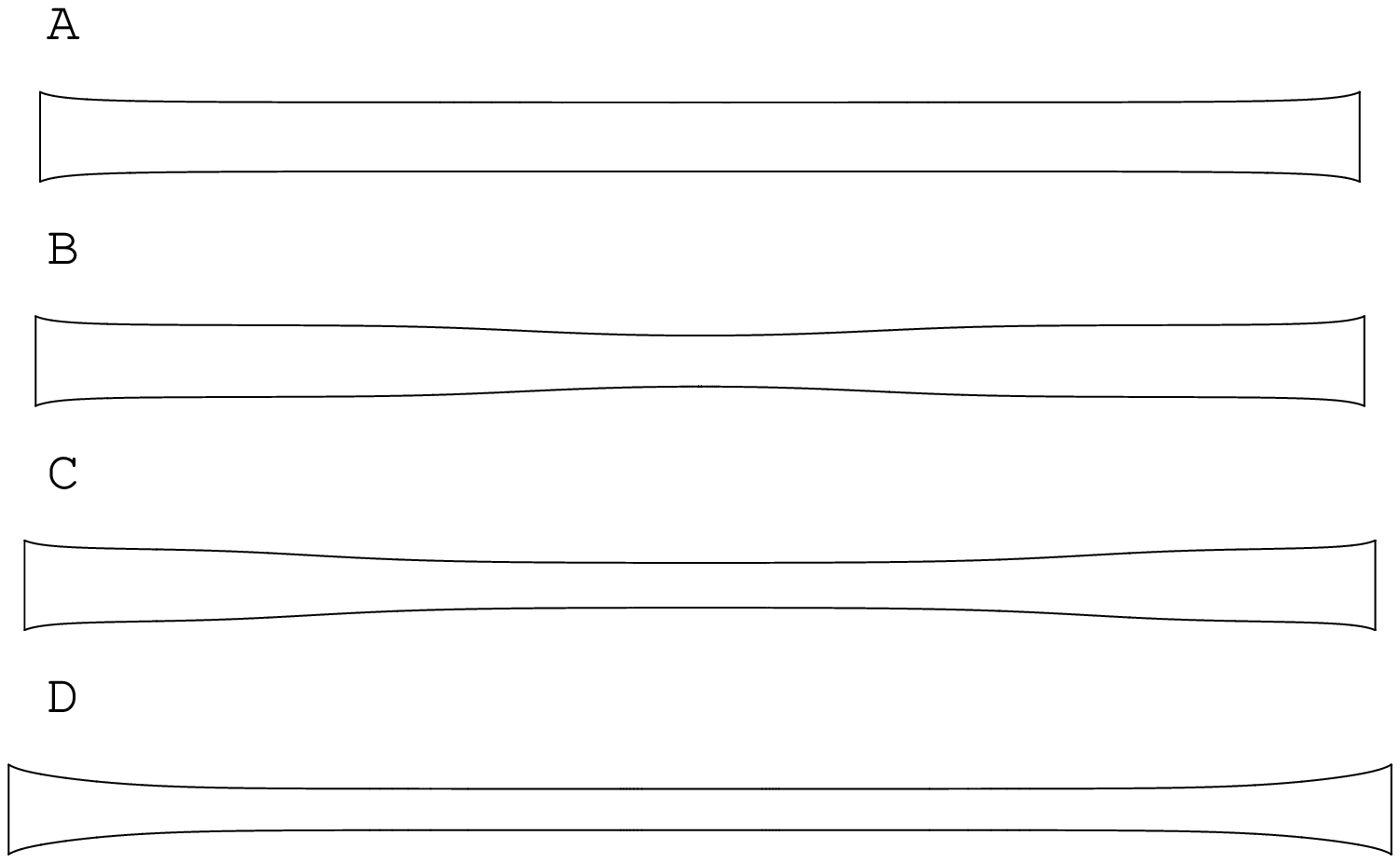}\\
\small\textcolor[rgb]{0.00,0.00,0.00}{Figure 8. Cylinder shapes in
different total elongations for the case $a=0.04$: (a)
$\Delta=0.035882224$; (b) $\Delta=0.043065200$; (c)
$\Delta=0.060283469$; (d) $\Delta=0.085396683$.}
\end{center}
Figure 8(a) represents a low-strain state, Figure 8(b) represents a
transition state, Figure 8(c) represents a two-phase state, and
Figure 8(d) represents a high-strain state. As the end displacement
increases these four states appears consecutively. Here, the
asymptotic solutions we obtained can describe the whole deformation
process. We know that the dimensionless engineering stress $\gamma$
represent the stress density acting on the cross section of the
cylinder in the reference configuration. Now, based on the solutions
obtained, we can also calculate the true stress density $\sigma$
acting on the cross section of the cylinder in the current
configuration. Figure 9 shows the distributions of the true stress
$\sigma$ corresponding to the four points $A-D$. We see, very
interesting, the true stress actually decreases from A to B and to
C. We also point out that a pure one-dimensional model cannot yield
the true stress distribution.
\begin{center}
\includegraphics[width=86mm,height=48mm]{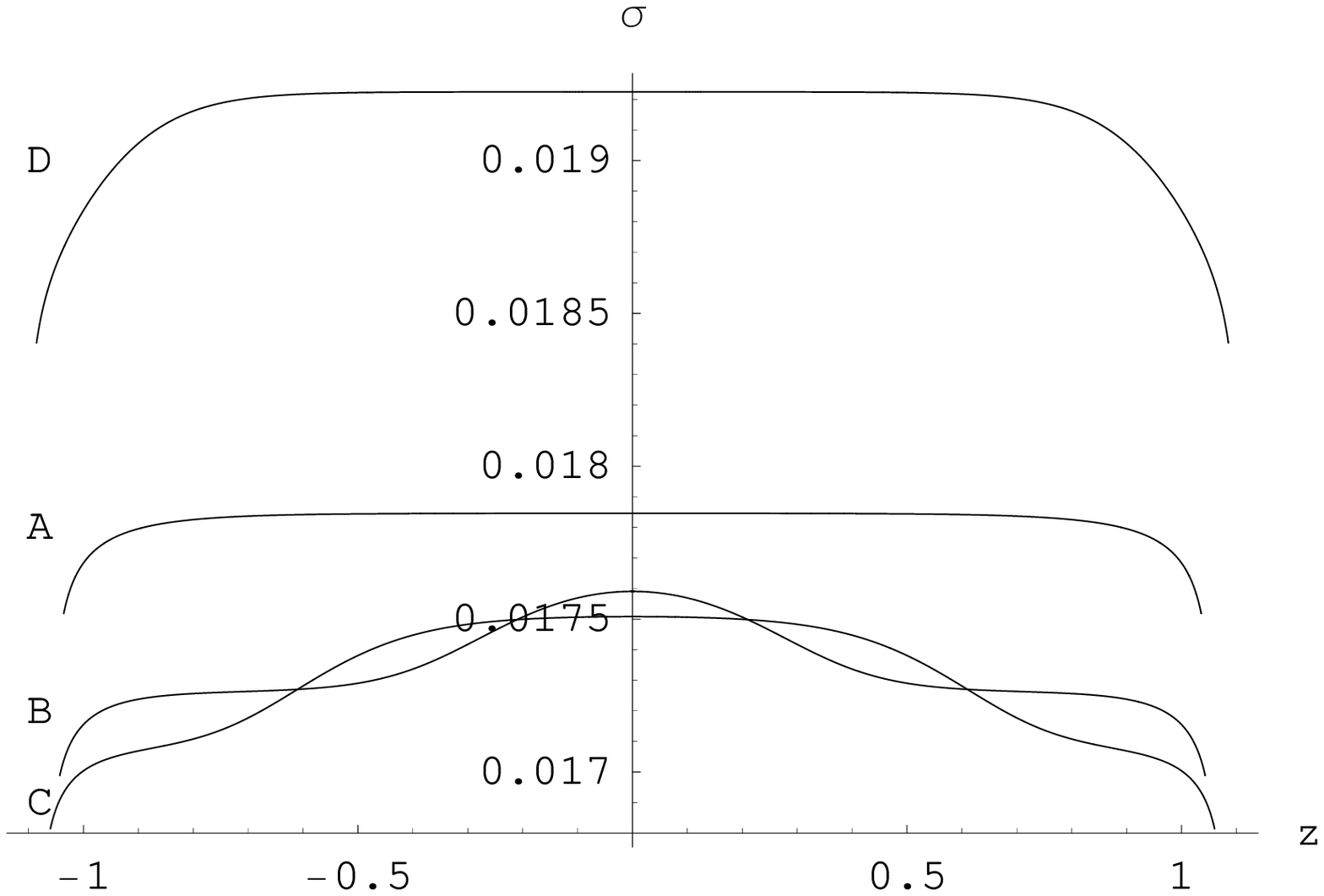}\\
\small\textcolor[rgb]{0.00,0.00,0.00}{Figure 9. The distributions of
the true stress density $\sigma$ acting on the cross section of the
cylinder in the current configuration.}
\end{center}

If the diameter-length ratio $a$ is relatively small, we know that
there is a snap-back on the corresponding $\gamma-\Delta$ curve (cf.
Figure 6(a) and 6(b)). Thus, for the total elongation $\Delta$
located in some special region, there may exist multiple
corresponding stress values $\gamma$, i.e., the solutions are not
unique. Then we need to determine which solution is the preferred
one.

We take $a=0.02$ as an example. From Figure 10, we can see that when
$\Delta_1<\Delta<\Delta_2$, there are three possible solutions,
which are labeled as $U_1$, $U_2$ and $U_3$, respectively.

\begin{center}
\includegraphics[width=80mm,height=50mm]{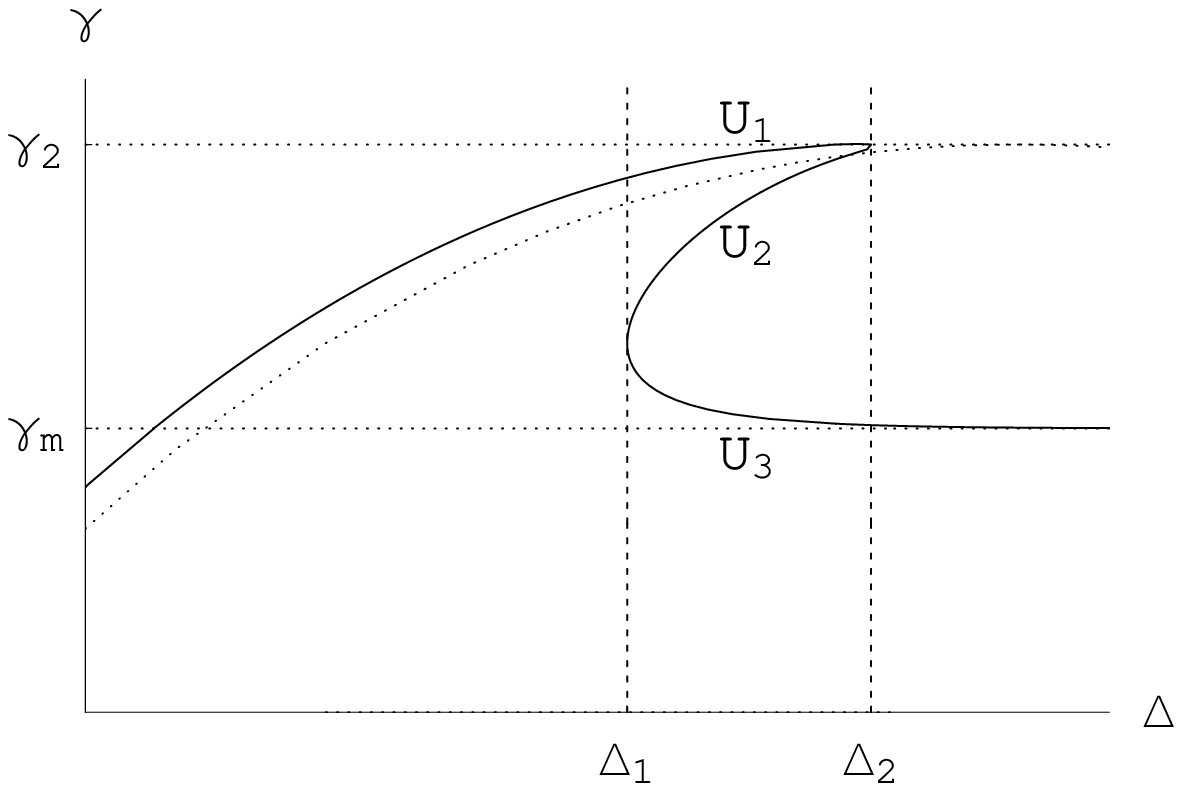}\\
\small\textcolor[rgb]{0.00,0.00,0.00}{Figure 10. Snap-back on the
$\gamma-\Delta$ curve for $a=0.02$.}
\end{center}
To choose the preferred solution from $U_1$, $U_2$ and $U_3$, we
calculate the total potential energy values of these solutions.
Denote $\Omega^{*}$ the total potential energy of the cylinder in
the displacement-controlled problem and we have
$$
\begin{aligned}
\Omega^* =2\pi a^2 E
\int_0^1&(\frac{1}{2}V^2+\frac{1}{3}D_1V^3+\frac{1}{4}D_2V^4-\frac{1}{8}a^2
VV_{ZZ}\\&+a^2(H_1VV_Z^2+H_2V^2V_{ZZ})) dZ.
\end{aligned}
\eqno(7.9)
$$
Here we still choose $H_1=15$, $H_2=5$. In Figure 11, we plot the
differences of the total potential energies between solutions $U_2$,
$U_3$ and solution $U_1$ for $\Delta_1<\Delta<\Delta_2$.

\begin{center}
\includegraphics[width=90mm,height=55mm]{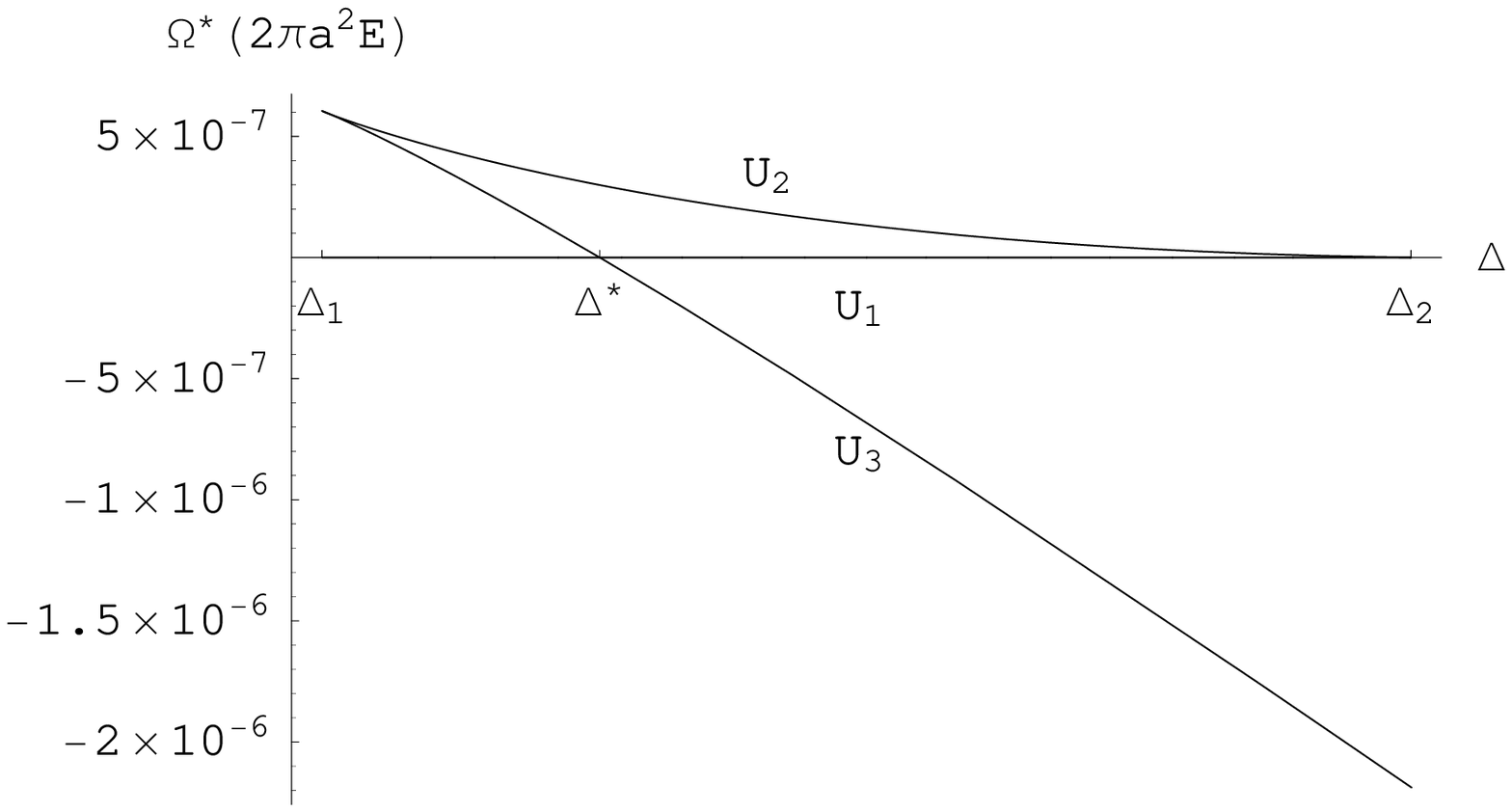}\\
\small\textcolor[rgb]{0.00,0.00,0.00}{Figure 11. Differences of the
total potential energies between solutions $U_2$, $U_3$ and solution
$U_1$ for $\Delta_1<\Delta<\Delta_2$ ($a=0.02$).}
\end{center}

From Figure 11, we can see that for
$\Delta_1<\Delta<\Delta^*(=0.038848990)$, solution $U_1$ is the
preferred solution and for $\Delta^*<\Delta<\Delta_2$, solution
$U_3$ is the preferred solution. The engineering stress-strain curve
corresponding to the preferred solution is shown in Figure 12.
\begin{center}
\includegraphics[width=90mm,height=55mm]{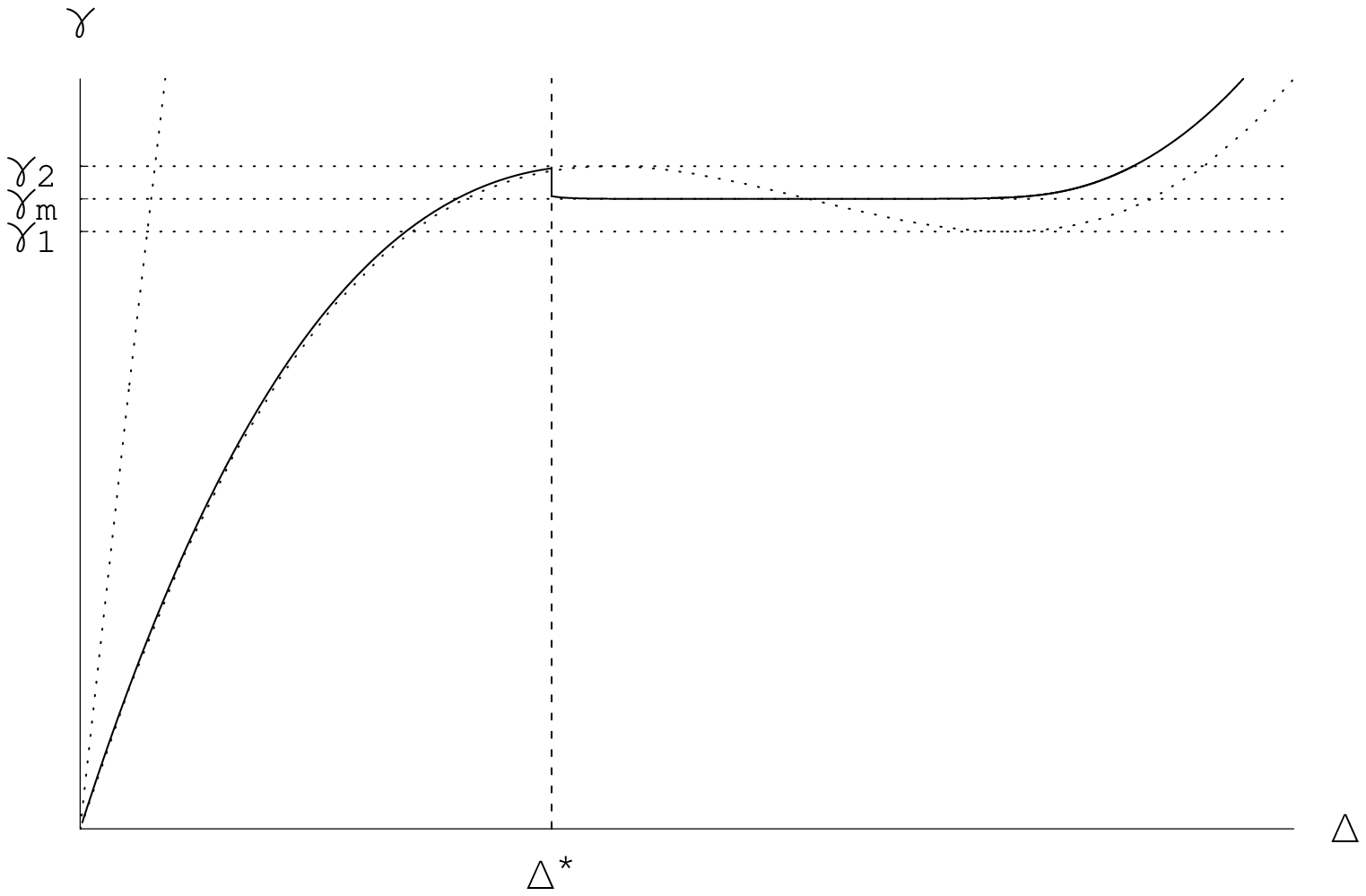}\\
\small\textcolor[rgb]{0.00,0.00,0.00}{Figure 12. The engineering
stress-strain curve corresponding to the preferred solution for
$a=0.02$.}
\end{center}
It is worth to note that there exists a jump for the stress $\gamma$
at $\Delta=\Delta^*$. This curve also captures the main features of
the experimental results in the loading process (Shaw $\&$
Kyriakides 1995, Sun \emph{et al.} 2000, Tse $\&$ Sun 2000, Favier
\emph{et al.} 2001 and Li $\&$ Sun 2002).

Figure 13 shows the true stress-strain ($\sigma$-$\Delta$) curves
(occurring at $Z=0$ and $Z=0.6$) corresponding to the preferred
solutions for $a=0.02$.
\begin{center}
$$
\begin{aligned}
\includegraphics[width=63mm,height=43mm]{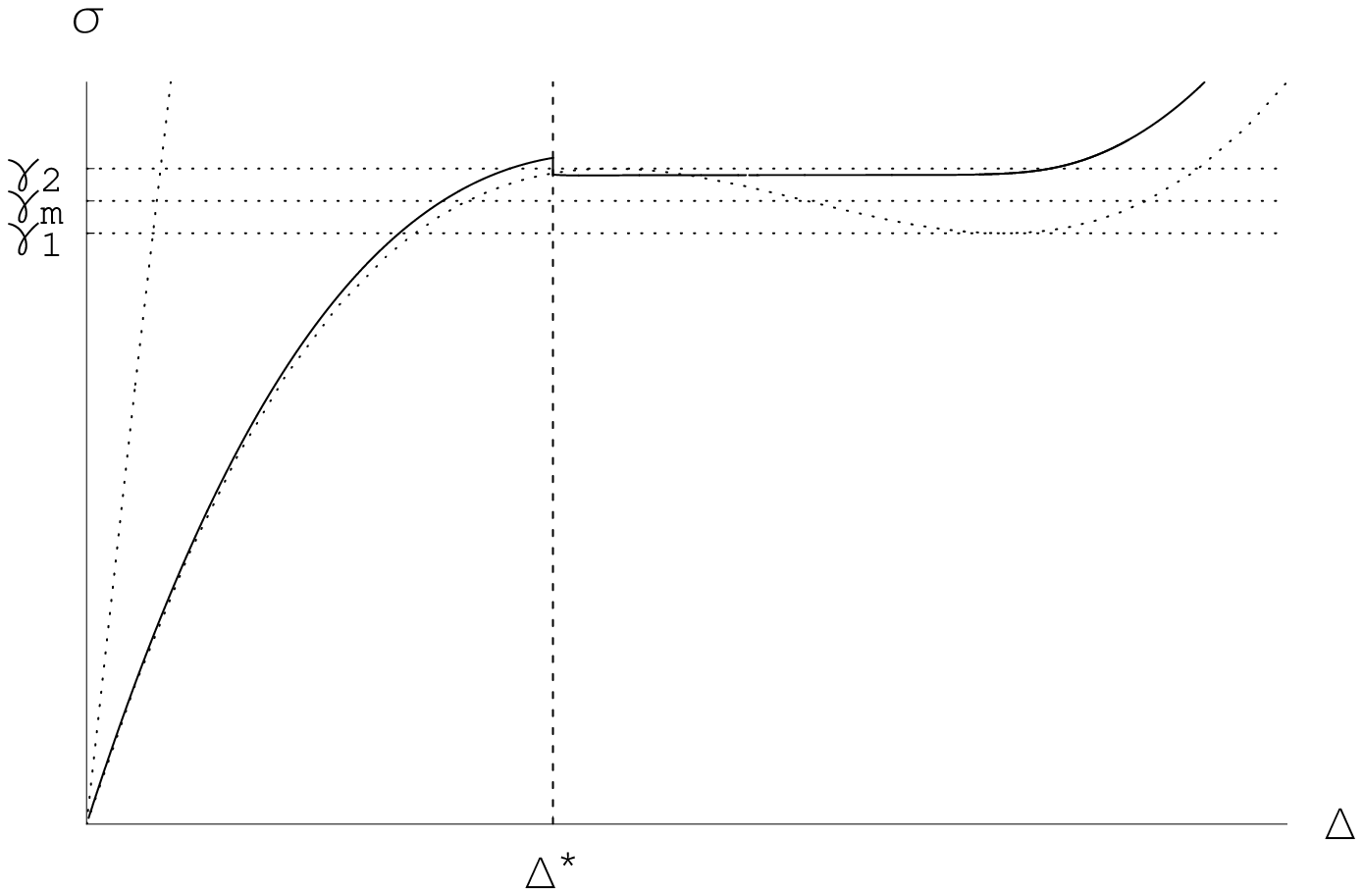}\\
\small\textcolor[rgb]{0.00,0.00,0.00}{(a)}
\end{aligned}\ \ \ \ \ \ \ \ \
\begin{aligned}
\includegraphics[width=63mm,height=43mm]{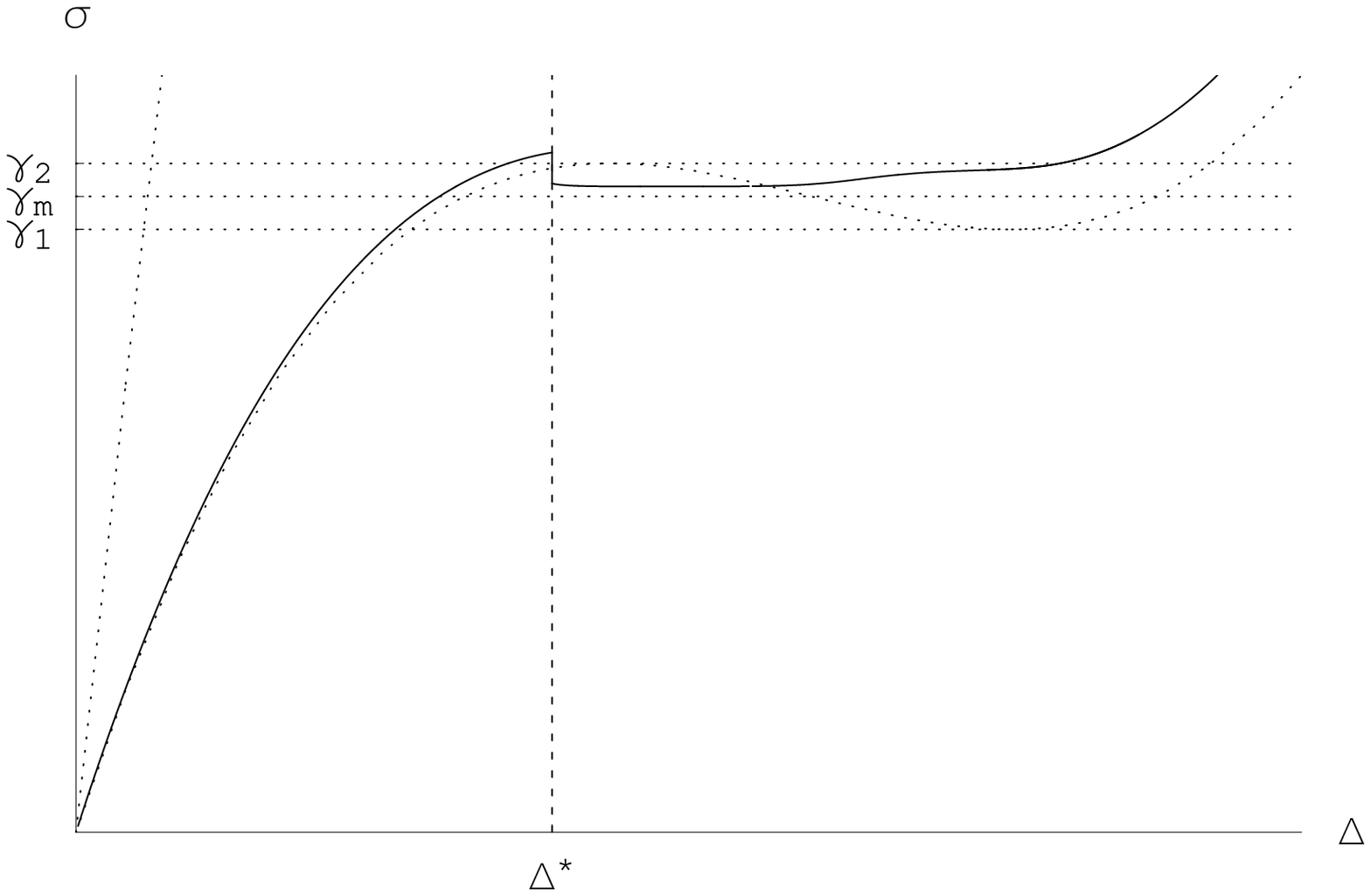}\\
\small\textcolor[rgb]{0.00,0.00,0.00}{(b)}
\end{aligned}
$$
\\
\small\textcolor[rgb]{0.00,0.00,0.00}{Figure 13. The true
stress-strain curves. (a) occurring at $Z=0$ for $a=0.02$; (b)
occurring at $Z=0.6$ for $a=0.02$.}
\end{center}
It is worth to note that for the case occurring at $Z=0$, the drop
for $\sigma$ become very small and after that there is also a stress
plateau (with the true stress value larger than the Maxwell stress).
For the case occurring at $Z=0.6$, the drops for $\sigma$ become
larger. But now the stress plateaus have been divided into two
parts, which corresponding to the fact that phase transition first
start at the middle part of the cylinder and transform gradually
towards the two ends.

%%%%%%%%%%%%%%%%%%%%%%%%%%%%%%%%%%%%%%%%%%%%%%%%%%%%%%%%%%%%%%%%%%%%%%%%%%%%%%%%%%%%%%%%%%%%%%%%%%%%%%%%%%%

%%%%%%%%%%%%%%%%%%%%%%%%%%%%%%%%%%%%%%%%%%%%%%%%%%%%%%%%%%%%%%%%%%%%%%%%%%%%%%%%%%%%%

\section{Conclusions}

We consider phase transitions induced by axial tension/extension in
a slender circular cylinder composed of a general compressible
hyperelastic material with a non-convex strain energy function under
clamped end conditions. In order to capture the macroscopic
phenomena observed in the experiments by others, the problem is
formulated in a three-dimensional setting and as a result it is
governed by a system of coupled nonlinear PDE's with complicated
nonlinear boundary conditions. A novel approach involving coupled
series-asymptotic expansions is developed to derive the normal form
equation (NFE) of the original system. By conducting the phase-plane
analysis on this NFE, we manage to deduce the global bifurcation
properties of its solutions. The solutions (including those for
post-bifurcation solutions) for both a force-controlled problem and
a displacement-controlled problem are obtained. These asymptotic
solutions demonstrate the essential features of phase transitions in
a cylinder and are consistent with the experimental data
qualitatively. Specifically, the engineering stress-strain curve is
shown to have the same features as observed in experiments. These
qualitative agreements with experiments give strong supporting
evidence that a non-convex strain energy function can be used to
describe phase transitions and the non-convexity of the strain
energy function is the main cause of the instability phenomena
associated with phase transitions in a slender cylinder. An
important finding is that under the clamped end conditions, there
exists a critical stress value $\gamma_0(a)$, which depends on the
radius-length ratio and larger than the Maxwell stress, such that
phase transition can only happen when the engineering stress is
larger than this value. This is different from the case of the
nature boundary condition that phase transition can happen as long
as the engineering stress is larger than the Maxwell stress.

%%%%%%%%%%%%%%%%%%%%%%%%%%%%%%%%%%%%%%%%%%%%%%%%%%%%%%%%%%%%%%%%%%%%%%%%%%%%%%%%%%%%%%%%%%%%%%%%%%%%%%%%%%%

\begin{flushleft}
{\bfseries\Large Acknowledgments}
\end{flushleft}

The work described in this paper was fully supported by two grants
from the Research Grants Council of the HKSAR, China (Project
numbers: CityU 100804 and CityU 100807).

%%%%%%%%%%%%%%%%%%%%%%%%%%%%%%%%%%%%%%%%%%%%%%%%%%%%%%%%%%%%%%%%%%%%%%%%%%%%%%%%%%%%%%%%%%%%%%%%%%%%%%%%%%%
\makeatletter
\def\app@number#1{ \setcounter{#1}{0}%
\@addtoreset{#1}{section}%
\@namedef{the#1}{\thesection.\arabic{#1}}}
\def\appendix{\@ifstar{\appendix@star}{\appendix@nostar}}
\def\appendix@nostar{%
\def\lb@section{ \appendixname \ \thesection.\half@em}
\def\lb@empty@section{\appendixname\ \thesection}
\setcounter{section}{0}\def\thesection{Appendix \Alph{section}}%
\setcounter{subsection}{0}%
\setcounter{subsubsection}{0}%
\setcounter{paragraph}{0}%
\app@number{equation}\app@number{figure}\app@number{table}}
\def\appendix@star{%
\def\lb@section{\appendixname}\let\lb@empty@section\lb@section
\setcounter{section}{0}\def\thesection{Appendix \Alph{section}}%
\setcounter{subsection}{0}%
\setcounter{subsubsection}{0}%
\setcounter{paragraph}{0}%
\app@number{equation}\app@number{figure}\app@number{table}}
\makeatother

\appendix

\section{Incremental elastic moduli}

For initially isotropic material, in the case that there are no
prestresses, $\Phi$ should be a function of the principle stretches
$\lambda_1$, $\lambda_2$ and $\lambda_3$, namely
$\Phi$=$\Phi(\lambda_1,\lambda_2,\lambda_3)$. Denote by
$\Phi_j=\frac{\partial \Phi}{\partial
\lambda_j}|_{\lambda_1=\lambda_2=\lambda_3=1}$, then
$\Phi_1=\Phi_2=\Phi_3$ should vanish since there are no prestresses.

The non-zero first order incremental elastic moduli can be written
as
$$
\begin{aligned}
&\xi_1=a^1_{1111}=\Phi_{11},\\
&\xi_2=a^1_{1122}=\Phi_{12},\\
&\xi_3=a^1_{1212}=\frac{1}{2}(\xi_1-\xi_2),\\
&\xi_4=a^1_{1221}=\xi_3.
\end{aligned}
$$
There are only two independent constants among $\xi_i$.

The non-zero second order incremental elastic moduli can be written
as
$$
\begin{aligned}
&\eta_1=a^2_{111111}=\Phi_{111},\\
&\eta_2=a^2_{111122}=\Phi_{112},\\
&\eta_3=a^2_{112233}=\Phi_{123},\\
&\eta_4=a^2_{111212}=\frac{1}{4}(2\xi_2+2\xi_3+\eta_1-\eta_2),\\
&\eta_5=a^2_{331212}=\frac{1}{2}(\xi_2+\eta_2-\eta_3),\\
&\eta_6=a^2_{121323}=\frac{1}{2}(\eta_4-\eta_5),\\
&\eta_7=a^2_{111221}=\eta_4-\xi_2-\xi_3,\\
&\eta_8=a^2_{331221}=\eta_5-\xi_2,\\
&\eta_9=a^2_{123123}=\eta_6-\xi_3.
\end{aligned}
$$
There are only three additional independent constants among
$\eta_i$.

The non-zero third order incremental elastic moduli can be written
as
$$
\begin{aligned}
&\theta_1=a^3_{11111111}=\Phi_{1111},\\
&\theta_2=a^3_{11111122}=\Phi_{1112},\\
&\theta_3=a^3_{11112222}=\Phi_{1122},\\
&\theta_4=a^3_{11112233}=\Phi_{1123},\\
&\theta_5=a^3_{11111212}=-\frac{1}{12}(6\xi_2+6\xi_3-3\eta_1-3\eta_2-2\theta_1+2\theta_2),\\
&\theta_6=a^3_{11112323}=\frac{1}{2}(\eta_2+\theta_3-\theta_4),\\
&\theta_7=a^3_{11221212}=-\frac{1}{2}(6\xi_2+6\xi_3-6\eta_2-\theta_1-2\theta_2+3\theta_3),\\
&\theta_8=a^3_{11221313}=-\frac{1}{4}(\xi_2-\eta_2-\eta_3-\theta_2+\theta_4),\\
&\theta_9=a^3_{12121212}=\frac{1}{8}(6\xi_2+6\xi_3+6\eta_1-6\eta_2+\theta_1-4\theta_2+3\theta_3),\\
\end{aligned}
$$
$$
\begin{aligned}
&\theta_{10}=a^3_{12121313}=\frac{1}{3}\theta_9,\\&\theta_{11}=a^3_{11121323}=-\frac{1}{24}(3\xi_1-3\eta_2+3\eta_3-\theta_1+\theta_2+3\theta_3-3\theta_4),\\
&\theta_{12}=a^3_{11111221}=\frac{1}{12}(6\xi_2+6\xi_3-3\eta_1-3\eta_2+2\theta_1-2\theta_2),\\
&\theta_{13}=a^3_{11112332}=-\frac{1}{2}(\eta_2-\theta_3+\theta_4),\\
&\theta_{14}=a^3_{11221221}=\frac{1}{12}(6\xi_2+6\xi_3-6\eta_2+\theta_1+2\theta_2-3\theta_3),\\
&\theta_{15}=a^3_{11221331}=\frac{1}{4}(\xi_2-\eta_2-\eta_3+\theta_2-\theta_4),\\
&\theta_{16}=a^3_{12121221}=-\frac{1}{8}(6\xi_2+6\xi_3-\theta_1+4\theta_2-3\theta_3),\\
&\theta_{17}=a^3_{12211221}=\frac{1}{8}(6\xi_2+6\xi_3-2\eta_1+2\eta_2+\theta_1-4\theta_2+3\theta_3),\\
&\theta_{18}=a^3_{12121331}=\frac{1}{3}\theta_{16},\\
&\theta_{19}=a^3_{12211331}=\frac{1}{24}(6\xi_1-3\eta_1-3\eta_2+6\eta_3+\theta_1-4\theta_2+3\theta_3),\\
&\theta_{20}=a^3_{11123123}=\frac{1}{24}(6\xi_1+3\xi_2-3\eta_1+3\eta_3+\theta_1-\theta_2-3\theta_3+3\theta_4),\\
&\theta_{21}=a^3_{11123132}=\theta_{20}-\frac{1}{4}(\xi_1+\xi_2-\eta_1+\eta_2),\\
&\theta_{22}=a^3_{12122323}=\frac{1}{24}(6\xi_2-6\xi_3+12\eta_2-12\eta_3+\theta_1-4\theta_2+3\theta_3).
\end{aligned}
$$
There are only four additional independent constants among
$\theta_i$.

The expressions of the constants $c_i$ ($i=1,\cdots,12$) in equation
(5.3) are given below:
$$
\begin{aligned}
c_1=&\frac{\eta_2}{4}+\frac{\xi_2\eta_2}{4\xi_1}+\frac{\xi_2\eta_2}{4\xi_3}+\frac{\xi_2^2\eta_2}{4\xi_1\xi_3}+\frac{\eta_3}{4}+\frac{\xi_2\eta_3}{4\xi_1}+\frac{\xi_2\eta_3}{4\xi_3}+\frac{\xi_2^2\eta_3}{4\xi_1\xi_3}-\frac{3\eta_4}{4}-\frac{5\xi_2\eta_4}{4\xi_1}\\
&-\frac{3\xi_2^2\eta_4}{8\xi_3^2}-\frac{5\xi_2^3\eta_4}{8\xi_1\xi_3^2}-\frac{3\xi_2\eta_4}{4\xi_3}-\frac{5\xi_2^2\eta_4}{4\xi_1\xi_3}+\frac{\eta_5}{4}-\frac{\xi_2\eta_5}{4\xi_1}+\frac{\xi_2^2\eta_5}{8\xi_3^2}-\frac{\xi_2^3\eta_5}{8\xi_1\xi_3^2}+\frac{\xi_2\eta_5}{4\xi_3}\\
&-\frac{\xi_2^2\eta_5}{4\xi_1\xi_3}+\frac{3\eta_7}{4}+\frac{5\xi_2\eta_7}{4\xi_1}+\frac{3\xi_2\eta_7}{4\xi_3}+\frac{5\xi_2^2\eta_7}{4\xi_1\xi_3}-\frac{\eta_8}{4}+\frac{\xi_2\eta_8}{4\xi_1}-\frac{\xi_2\eta_8}{4\xi_3}+\frac{\xi_2^2\eta_8}{4\xi_1\xi_3},\\
c_2=&\frac{\xi_2\eta_2}{4\xi_1}-\frac{\xi_1\eta_2}{4\xi_3}+\frac{\xi_2\eta_2}{2\xi_3}+\frac{\xi_2^2\eta_2}{4\xi_1\xi_3}+\frac{\xi_2\eta_3}{4\xi_1}-\frac{\xi_1\eta_3}{4\xi_3}+\frac{\xi_2^2\eta_3}{4\xi_1\xi_3}+\frac{\eta_4}{2}-\frac{5\xi_2\eta_4}{4\xi_1}+\frac{\xi_1\xi_2\eta_4}{4\xi_3^2}\\
&-\frac{\xi_2^2\eta_4}{4\xi_3^2}-\frac{5\xi_2^3\eta_4}{8\xi_1\xi_3^2}+\frac{\xi_1\eta_4}{4\xi_3}-\frac{5\xi_2^2\eta_4}{4\xi_1\xi_3}-\frac{\xi_2\eta_5}{4\xi_1}+\frac{\xi_1\xi_2\eta_5}{4\xi_3^2}-\frac{\xi_2^3\eta_5}{8\xi_1\xi_3^2}+\frac{\xi_1\eta_5}{4\xi_3}-\frac{\xi_2^2\eta_5}{4\xi_1\xi_3}-\frac{\eta_7}{2}\\
&+\frac{5\xi_2\eta_7}{4\xi_1}-\frac{\xi_1\eta_7}{4\xi_3}+\frac{5\xi_2^2\eta_7}{4\xi_1\xi_3}+\frac{\xi_2\eta_8}{4\xi_1}-\frac{\xi_1\eta_8}{4\xi_3}+\frac{\xi_2^2\eta_8}{4\xi_1\xi_3},\\
\end{aligned}
$$
$$
\begin{aligned}
c_3=&-\frac{\xi_2\eta_1}{4\xi_1}-\frac{\xi_2^2\eta_1}{2\xi_1^2}-\frac{\xi_2^2\eta_1}{8\xi_1\xi_3}-\frac{\xi_2^3\eta_1}{4\xi_1^2\xi_3}+\frac{\eta_2}{4}+\frac{3\xi_2\eta_2}{4\xi_1}-\frac{\xi_2^2\eta_2}{2\xi_1^2}-\frac{\xi_2\eta_2}{4\xi_3}+\frac{5\xi_2^2\eta_2}{8\xi_1\xi_3}-\frac{\xi_2^3\eta_2}{4\xi_1^2\xi_3}\\
&+\frac{\eta_3}{4}+\frac{\xi_2\eta_3}{2\xi_1}-\frac{\xi_2\eta_3}{4\xi_3}+\frac{\xi_2^2\eta_3}{2\xi_1\xi_3}-\frac{\eta_4}{2}-\frac{\xi_2\eta_4}{2\xi_1}+\frac{\xi_2^2\eta_4}{4\xi_3^2}-\frac{\xi_2^3\eta_4}{4\xi_1\xi_3^2}-\frac{\xi_2^2\eta_4}{2\xi_1\xi_3}-\frac{\eta_5}{2}\\
&-\frac{\xi_2\eta_5}{2\xi_1}+\frac{\xi_2^2\eta_5}{4\xi_3^2}-\frac{\xi_2^3\eta_5}{4\xi_1\xi_3^2}-\frac{\xi_2^2\eta_5}{2\xi_1\xi_3}+\frac{\eta_7}{2}+\frac{\xi_2\eta_7}{2\xi_1}+\frac{\xi_2^2\eta_7}{2\xi_1\xi_3}+\frac{\eta_8}{2}+\frac{\xi_2\eta_8}{2\xi_1}+\frac{\xi_2^2\eta_8}{2\xi_1\xi_3},\\
c_4=&-\frac{\xi_2^2\eta_1}{2\xi_1^2}-\frac{\xi_2^3\eta_1}{4\xi_1^2\xi_3}+\frac{\xi_2\eta_2}{\xi_1}-\frac{\xi_2^2\eta_2}{\xi_1^2}-\frac{\xi_1\eta_2}{4\xi_3}-\frac{\xi_2\eta_2}{2\xi_3}+\frac{\xi_2^2\eta_2}{\xi_1\xi_3}-\frac{\xi_2^3\eta_2}{2\xi_1^2\xi_3}+\frac{\xi_2\eta_3}{\xi_1}-\frac{\xi_1\eta_3}{4\xi_3}\\
&+\frac{3\xi_2^2\eta_3}{4\xi_1\xi_3}-\frac{\eta_4}{2}-\frac{\xi_2\eta_4}{\xi_1}+\frac{\xi_1\xi_2\eta_4}{4\xi_3^2}+\frac{\xi_2^2\eta_4}{4\xi_3^2}-\frac{\xi_2^3\eta_4}{2\xi_1\xi_3^2}+\frac{\xi_1\eta_4}{4\xi_3}-\frac{\xi_2^2\eta_4}{\xi_1\xi_3}-\frac{\xi_2\eta_5}{2\xi_1}+\frac{\xi_1\xi_2\eta_5}{4\xi_3^2}\\
&-\frac{\xi_2^3\eta_5}{4\xi_1\xi_3^2}+\frac{\xi_1\eta_5}{4\xi_3}-\frac{\xi_2^2\eta_5}{2\xi_1\xi_3}+\frac{\eta_7}{2}+\frac{\xi_2\eta_7}{\xi_1}-\frac{\xi_1\eta_7}{4\xi_3}+\frac{\xi_2^2\eta_7}{\xi_1\xi_3}+\frac{\xi_2\eta_8}{2\xi_1}-\frac{\xi_1\eta_8}{4\xi_3}+\frac{\xi_2^2\eta_8}{2\xi_1\xi_3},\\
c_5=&-\frac{\eta_1}{8}-\frac{\xi_2\eta_1}{8\xi_3}+\frac{3\xi_2\eta_2}{4\xi_1}-\frac{\xi_2^2\eta_2}{2\xi_1^2}-\frac{\xi_1\eta_2}{4\xi_3}+\frac{5\xi_2^2\eta_2}{8\xi_1\xi_3}-\frac{\xi_2^3\eta_2}{4\xi_1^2\xi_3}-\frac{\xi_2\eta_4}{2\xi_1}+\frac{\xi_1\xi_2\eta_4}{4\xi_3^2}\\
&-\frac{\xi_2^3\eta_4}{4\xi_1\xi_3^2}+\frac{\xi_1\eta_4}{4\xi_3}-\frac{\xi_2^2\eta_4}{2\xi_1\xi_3}+\frac{\xi_2\eta_7}{2\xi_1}-\frac{\xi_1\eta_7}{4\xi_3}+\frac{\xi_2^2\eta_7}{2\xi_1\xi_3},\\
c_6=&\frac{\eta_2}{2}+\frac{\xi_2\eta_2}{2\xi_1}+\frac{\xi_1\eta_2}{4\xi_3}+\frac{\xi_2^2\eta_2}{2\xi_1\xi_3}+\frac{\xi_1\eta_3}{4\xi_3}-\frac{\eta_4}{2}-\frac{\xi_2\eta_4}{2\xi_1}-\frac{\xi_1\xi_2\eta_4}{2\xi_3^2}-\frac{\xi_2^2\eta_4}{4\xi_3^2}-\frac{\xi_2^3\eta_4}{4\xi_1\xi_3^2}\\
&-\frac{\xi_1\eta_4}{2\xi_3}-\frac{\xi_2\eta_4}{2\xi_3}-\frac{\xi_2^2\eta_4}{2\xi_1\xi_3}+\frac{\eta_7}{2}+\frac{\xi_2\eta_7}{2\xi_1}+\frac{\xi_1\eta_7}{2\xi_3}+\frac{\xi_2\eta_7}{2\xi_3}+\frac{\xi_2^2\eta_7}{2\xi_1\xi_3},\\
c_7=&\frac{\xi_2\eta_1}{4\xi_3}+\frac{\xi_2\eta_2}{2\xi_1}-\frac{\xi_1\eta_2}{4\xi_3}+\frac{\xi_2^2\eta_2}{2\xi_1\xi_3}-\frac{\xi_2\eta_4}{2\xi_1}+\frac{\xi_1^2\eta_4}{8\xi_3^2}-\frac{\xi_2^2\eta_4}{2\xi_3^2}-\frac{\xi_2^3\eta_4}{4\xi_1\xi_3^2}+\frac{\xi_1\eta_4}{4\xi_3}-\frac{\xi_2\eta_4}{2\xi_3}\\
&-\frac{\xi_2^2\eta_4}{2\xi_1\xi_3}+\frac{\xi_1^2\eta_5}{8\xi_3^2}+\frac{\xi_2\eta_7}{2\xi_1}-\frac{\xi_1\eta_7}{4\xi_3}+\frac{\xi_2\eta_7}{2\xi_3}+\frac{\xi_2^2\eta_7}{2\xi_1\xi_3},\\
c_8=&\frac{\eta_1}{8}+\frac{\xi_2\eta_1}{8\xi_1}-\frac{\xi_2\eta_1}{8\xi_3}+\frac{\xi_2^2\eta_1}{8\xi_1\xi_3}+\frac{\xi_1\eta_2}{4\xi_3}-\frac{5\xi_1^2\eta_4}{32\xi_3^2}+\frac{\xi_1\xi_2\eta_4}{32\xi_3^2}-\frac{\xi_2^2\eta_4}{8\xi_3^2}-\frac{\xi_1\eta_4}{8\xi_3}-\frac{\xi_2\eta_4}{8\xi_3}\\
&-\frac{\xi_1^2\eta_5}{32\xi_3^2}+\frac{\xi_1\xi_2\eta_5}{32\xi_3^2}+\frac{\xi_1\eta_7}{8\xi_3}+\frac{\xi_2\eta_7}{8\xi_3},\\
c_9=&\frac{\xi_2\eta_1}{8\xi_1}+\frac{\xi_2^2\eta_1}{8\xi_1\xi_3}+\frac{\xi_1^2\eta_4}{16\xi_3^2}-\frac{3\xi_1\xi_2\eta_4}{32\xi_3^2}-\frac{\xi_2^2\eta_4}{8\xi_3^2}-\frac{\xi_2\eta_4}{8\xi_3}+\frac{\xi_1\xi_2\eta_5}{32\xi_3^2}+\frac{\xi_2\eta_7}{8\xi_3},\\
c_{10}=&-\frac{\eta_1}{16}-\frac{\xi_2\eta_1}{8\xi_1}-\frac{\xi_2\eta_1}{16\xi_3}-\frac{\xi_2^2\eta_1}{8\xi_1\xi_3}+\frac{5\eta_2}{16}+\frac{\xi_2\eta_2}{8\xi_1}-\frac{\xi_2\eta_2}{16\xi_3}+\frac{\xi_2^2\eta_2}{8\xi_1\xi_3}+\frac{\xi_2\eta_3}{8\xi_3}+\frac{\xi_1\xi_2\eta_4}{8\xi_3^2}\\
&-\frac{\xi_2^2\eta_4}{8\xi_3^2}-\frac{\xi_1\eta_4}{8\xi_3}-\frac{\xi_2\eta_4}{8\xi_3}+\frac{\xi_1\xi_2\eta_5}{8\xi_3^2}-\frac{\xi_2^2\eta_5}{8\xi_3^2}-\frac{\xi_1\eta_5}{8\xi_3}-\frac{\xi_2\eta_5}{8\xi_3}+\frac{\xi_1\eta_7}{8\xi_3}+\frac{\xi_2\eta_7}{8\xi_3}\\
&+\frac{\xi_1\eta_8}{8\xi_3}+\frac{\xi_2\eta_8}{8\xi_3},\\
c_{11}=&\frac{\eta_1}{8}-\frac{\xi_2\eta_1}{8\xi_3}-\frac{3\xi_1\eta_2}{8\xi_3}+\frac{\xi_2\eta_2}{4\xi_3}+\frac{\eta_3}{8}+\frac{\xi_2\eta_3}{4\xi_3}+\frac{\xi_1^2\eta_4}{8\xi_3^2}+\frac{\xi_1\xi_2\eta_4}{8\xi_3^2}-\frac{\xi_2^2\eta_4}{4\xi_3^2}-\frac{\xi_1\eta_4}{8\xi_3}\\
&-\frac{\xi_2\eta_4}{4\xi_3}+\frac{\xi_1^2\eta_5}{8\xi_3^2}-\frac{\xi_2^2\eta_5}{8\xi_3^2}+\frac{\xi_2\eta_5}{8\xi_3}+\frac{\xi_1\eta_7}{8\xi_3}+\frac{\xi_2\eta_7}{4\xi_3}+\frac{\xi_2\eta_8}{8\xi_3},\\
\end{aligned}
$$
$$
\begin{aligned}
c_{12}=&\frac{\xi_2\eta_1}{8\xi_1}-\frac{3\xi_1\eta_1}{16\xi_3}+\frac{\xi_2^2\eta_1}{8\xi_1\xi_3}+\frac{\eta_2}{16}-\frac{\xi_2\eta_2}{8\xi_1}+\frac{3\xi_2\eta_2}{16\xi_3}-\frac{\xi_2^2\eta_2}{8\xi_1\xi_3}+\frac{\xi_1^2\eta_4}{8\xi_3^2}-\frac{\xi_2^2\eta_4}{8\xi_3^2}-\frac{\xi_2\eta_4}{8\xi_3}+\frac{\xi_2\eta_7}{8\xi_3}.
\end{aligned}
$$

\section{Non-dimensional field equations}

The full forms of the non-dimensional field equations (3.3) and
(3.4) are given below:
$$
\begin{aligned}
&(2\xi_2+2\xi_3)v_z+\xi_1w_{zz}+4\xi_3w_s+s((2\xi_2+2\xi_3)v_{sz}+4\xi_3w_{ss})\\
&+\epsilon[(2\eta_2+2\eta_3+2\eta_7+2\eta_8)vv_z+(2\eta_2+2\eta_7)v_zw_z+2\eta_2vw_{zz}+\eta_1w_zw_{zz}\\
&+(4\eta_4+4\eta_5)vw_s+4\eta_4w_zw_s+s(\eta_4v_zv_{zz}+(2\eta_2+2\eta_3+10\eta_7+2\eta_8)v_sv_z\\
&+2\eta_2w_{zz}v_s+2\eta_7v_{zz}w_s+(20\eta_4+4\eta_5)v_sw_s+(2\eta_2+2\eta_3+2\eta_7+2\eta_8)vv_{sz}\\
&+(2\eta_2+2\eta_7)w_sv_{sz}+4\eta_7v_zw_{sz}+8\eta_4w_sw_{sz}+(4\eta_4+4\eta_5)vw_{ss}+4\eta_4w_zw_{ss}\\
&+s^2((4\eta_2+4\eta_7)v_sv_{sz}+4\eta_7v_zv_{ss}+8\eta_4w_sv_{ss}+8\eta_4v_sw_{ss})]+\epsilon^2[(\theta_2+3\theta_4+\theta_{12}\\
&+\theta_{13}+2\theta_{15})v^2v_z+(2\theta_3+2\theta_4+2\theta_{14}+2\theta_{15})vv_zw_z+(\theta_2+\theta_{12})v_zw_z^2+(\theta_3+\theta_4)v^2w_{zz}\\
&+2\theta_2vw_zw_{zz}+\frac{1}{2}\theta_1w_z^2w_{zz}+(2\theta_5+2\theta_6+4\theta_8)v^2w_s+(4\theta_7+4\theta_8)vw_zw_s+2\theta_5w_z^2w_s\\
&+s((\frac{2}{3}\theta_{16}+\frac{1}{2}\theta_7+\frac{1}{2}\theta_8)v_z^3+(\theta_7+\theta_8)vv_zv_{zz}+\theta_5v_zw_zv_{zz}+\frac{1}{2}\theta_5v_z^2w_{zz}\\
&+(10\theta_{12}+2\theta_{13}+12\theta_{15}+2\theta_2+6\theta_4)vv_zv_s+(10\theta_{14}+2\theta_{15}+2\theta_3+2\theta_4)v_zw_zv_s\\
&+(2\theta_3+2\theta_4)vw_{zz}v_s+2\theta_2w_zw_{zz}v_s+(2\theta_{14}+2\theta_{15}+4\theta_{17})v_z^2w_s+(2\theta_{14}+2\theta_{15})vv_{zz}w_s\\
&+2\theta_{12}w_zv_{zz}w_s+2\theta_{12}v_zw_{zz}w_s+(20\theta_5+4\theta_6+24\theta_8)vv_sw_s+(20\theta_7+4\theta_8)w_zv_sw_s\\
&+(8\theta_{16}+2\theta_7+2\theta_8)v_zw_s^2+2\theta_5w_{zz}w_s^2+\frac{16}{3}\theta_9w_s^3+(\theta_{12}+\theta_{13}+2\theta_{15}+\theta_2+3\theta_4)v^2v_{sz}\\
&+(2\theta_{14}+2\theta_{15}+2\theta_3+2\theta_4)vw_zv_{sz}+(\theta_{12}+\theta_2)w_z^2v_{sz}+(4\theta_{14}+4\theta_{15})vv_zw_{sz}+2\theta_5w_z^2w_{ss}\\
&+4\theta_{12}v_zw_zw_{sz}+(8\theta_7+8\theta_8)vw_sw_{sz}+8\theta_5w_zw_sw_{sz}+(2\theta_5+2\theta_{6}+4\theta_8)v^2w_{ss}\\
&+(4\theta_7+4\theta_8)vw_zw_{ss})+s^2(2\theta_7v_zv_{zz}v_s+(16\theta_{12}+4\theta_{15}+2\theta_2+2\theta_4)v_zv_s^2+16\theta_7v_sw_sw_{sz}\\
&+2\theta_3w_{zz}v_s^2+4\theta_{14}v_{zz}v_sw_s+(\theta_{16}+\theta_7)v_z^2v_{sz}+(4\theta_{12}+4\theta_{15}+4\theta_2+4\theta_4)vv_sv_{sz}\\
&+(4\theta_{14}+4\theta_3)w_zv_sv_{sz}+(4\theta_{14}+4\theta_{17})v_zw_sv_{sz}+(4\theta_{16}+4\theta_7)w_s^2v_{sz}+8\theta_{14}v_zv_sw_{sz}\\
\end{aligned}
$$
$$
\begin{aligned}
&+(4\theta_{12}+4\theta_{15})vv_zv_{ss}+4\theta_{14}v_zw_zv_{ss}+(8\theta_5+8\theta_8)vw_sv_{ss}+8\theta_7w_zw_sv_{ss}+2\theta_{17}v_z^2w_{ss}\\
&+(8\theta_5+8\theta_8)vv_sw_{ss}+8\theta_7w_zv_sw_{ss}+8\theta_{16}v_zw_sw_{ss}+8\theta_9w_s^2w_{ss}+(32\theta_5+8\theta_8)v_s^2w_s)\\
&+s^3((4\theta_{12}+4\theta_2)v_s^2v_{sz}+8\theta_{12}v_zv_sv_{ss}+16\theta_5v_sw_sv_{zz}+8\theta_5v_s^2w_{ss})]=0,
\end{aligned}
\eqno(B1)
$$
$$
\begin{aligned}
&\xi_3v_{zz}+8\xi_1v_s+(2\xi_2+2\xi_3)w_{sz}+4s\xi_1v_{ss}+\epsilon[(\frac{5}{2}\eta_4+\frac{1}{2}\eta_5)v_z^2+(\eta_4+\eta_5)vv_{zz}\\
&+\eta_4w_zv_{zz}+\eta_4v_zw_{zz}+(8\eta_1+8\eta_2)vv_s+8\eta_2w_zv_s+8\eta_7v_zw_s+2\eta_7w_{zz}w_{s}\\
&+(6\eta_4-2\eta_5)w_s^2+(2\eta_2+2\eta_3+2\eta_7+2\eta_8)vw_{sz}+(2\eta_2+2\eta_7)w_zw_{sz}+s(2\eta_4v_{zz}v_s\\
&+(14\eta_1+2\eta_2)v_s^2+4\eta_4v_zv_{sz}+8\eta_7w_sv_{sz}+(4\eta_2+4\eta_7)v_sw_{sz}+(4\eta_1+4\eta_2)vv_{ss}\\
&+4\eta_2w_zv_{ss}+4\eta_7v_zw_{ss}+8\eta_4w_sw_{ss})+8s^2\eta_1v_sv_{ss}]+\epsilon^2[(\frac{5}{2}\theta_5+\frac{1}{2}\theta_6+3\theta_8)vv_z^2\\
&+(\frac{5}{2}\theta_7+\frac{1}{2}\theta_8)v_z^2w_z+(\frac{1}{2}\theta_5+\frac{1}{2}\theta_6+\theta_8)v^2v_{zz}+(\theta_7+\theta_8)vw_zv_{zz}+\frac{1}{2}\theta_5w_z^2v_{zz}\\
&+(\theta_7+\theta_8)vv_zw_{zz}+\theta_5v_zw_zw_{zz}+(4\theta_1+8\theta_2+4\theta_3)v^2v_s+(8\theta_2+8\theta_4)vw_zv_s\\
&+4\theta_3w_z^2v_s+(8\theta_{12}+8\theta_{15})vv_zw_s+8\theta_{14}v_zw_zw_s+(2\theta_{14}+2\theta_{15})vw_{zz}w_s+2\theta_{12}w_zw_{zz}w_s\\
&+(6\theta_5-2\theta_6+4\theta_8)vw_s^2+(6\theta_7-2\theta_8)w_zw_s^2+(\theta_{12}+\theta_{13}+2\theta_{15}+\theta_2+3\theta_4)v^2w_{sz}\\
&+(2\theta_{14}+2\theta_{15}+2\theta_3+2\theta_4)vw_zw_{sz}+(\theta_{12}+\theta_2)w_z^2w_{sz}+s(\frac{1}{2}\theta_9v_z^2v_{zz}+(8\theta_5+2\theta8)v_z^2v_s\\
&+(2\theta_5+2\theta_8)vv_{zz}v_s+2\theta_7w_zv_{zz}v_s+2\theta_7v_zw_{zz}v_s+(14\theta_1+16\theta_2+2\theta_3)vv_s^2\\
&+(14\theta_2+2\theta_4)w_zv_s^2+2\theta_{16}v_zv_{zz}w_s+(28\theta_{12}+4\theta_{15})v_zv_sw_s+4\theta_{14}w_{zz}v_sw_s\\
&+2\theta_{17}v_{zz}w_s^2+24\theta_5v_sw_s^2+(4\theta_5+4\theta_8)vv_zv_{sz}+4\theta_7v_zw_zv_{sz}+(8\theta_{12}+8\theta_{15})vw_sv_{sz}\\
&+8\theta_{14}w_zw_sv_{sz}+(\theta_{16}+\theta_7)v_z^2w_{sz}+(4\theta_{12}+4\theta_{15}+4\theta_2+4\theta_4)vv_sw_{sz}\\
&+(4\theta_{14}+4\theta_3)w_zv_sw_{sz}+(4\theta_{14}+4\theta_{17})v_zw_sw_{sz}+(4\theta_{16}+4\theta_7)w_s^2w_{sz}\\
&+(2\theta_1+4\theta_2+2\theta_3)v^2v_{ss}+(4\theta_2+4\theta_4)vw_zv_{ss}+2\theta_3w_z^2v_{ss}+(4\theta_{12}+4\theta_{15})vv_zw_{ss}\\
&+4\theta_{14}v_zw_zw_{ss}+(8\theta_5+8\theta_8)vw_sw_{ss}+8\theta_7w_zw_sw_{ss})+s^2(2\theta_5v_{zz}v_s^2+(\frac{40}{3}\theta_1+\frac{8}{3}\theta_2)v_s^3\\
&+8\theta_5v_zv_sv_{sz}+16\theta_{12}v_sw_sv_{sz}+(4\theta_{12}+4\theta_2)v_s^2w_{sz}+2\theta_5v_z^2v_{ss}+(8\theta_1+8\theta_2)vv_sv_{ss}\\
&+8\theta_2w_zv_sv_{ss}+8\theta_{12}v_zw_sv_{ss}+8\theta_5w_s^2v_{ss}+8\theta_{12}v_zv_sw_{ss}+16\theta_5v_sw_sw_{ss})\\
&+8s^3\theta_1v_s^2v_{ss}]=0.
\end{aligned}
\eqno(B2)
$$

\section{}

The full forms of the terms $H_1$-$H_7$ in equations (4.4)-(4.8) are
given below:
$$
\begin{aligned}
H_1=&(2\theta_5+2\theta_6+4\theta_8)V_0^2W_1+(\theta_{12}+\theta_{13}+2\theta_{15}+\theta_2+3\theta_4)V_0^2V_{0z}\\
&+(4\theta_7+4\theta_8)V_0W_1W_{0z}+(2\theta_{14}+2\theta_{15}+2\theta_3+2\theta_4)V_0V_{0z}W_{0z}+2\theta_5W_1W_{0z}^2\\
&+(\theta_{12}+\theta_2)V_{0z}W_{0z}^2+(\theta_3+\theta_4)V_0^2W_{0zz}+2\theta_2V_0W_{0z}W_{0zz}+\frac{1}{2}\theta_1W_{0z}^2W_{0zz},\\
H_2=&(24\theta_5+8\theta_6+32\theta_8)V_0V_1W_1+\frac{16}{3}\theta_9W_1^3+(8\theta_5+8\theta_6+16\theta_8)V_0^2W_2\\
&+(12\theta_{12}+4\theta_{13}+16\theta_{15}+4\theta_2+12\theta_4)V_0V_1V_{0z}+(8\theta_{16}+2\theta_7+2\theta_8)W_1^2V_{0z}\\
&+(2\theta_{14}+2\theta_{15}+4\theta_{17})W_1V_{0z}^2+(\frac{2}{3}\theta_{16}+\frac{1}{2}\theta_7+\frac{1}{2}\theta_8)V_{0z}^3+8\theta_5W_2W_{0z}^2\\
&+(2\theta_{12}+2\theta_{13}+4\theta_{15}+2\theta_2+6\theta_4)V_0^2V_{1z}+(24\theta_7+8\theta_8)V_1W_1W_{0z}\\
&+(16\theta_7+16\theta_8)V_0W_2W_{0z}+(12\theta_{14}+4\theta_{15}+4\theta_3+4\theta_4)V_1V_{0z}W_{0z}+(4\theta_{14}+4\theta_{15}\\
&+4\theta_3+4\theta_4)V_0V_{1z}W_{0z}+(12\theta_{12}+2\theta_2)V_{1z}W_{0z}^2+(12\theta_7+12\theta_8)V_0W_1W_{1z}\\
&+(6\theta_{14}+6\theta_{15}+2\theta_3+2\theta_4)V_0V_{0z}W_{1z}+12\theta_5W_1W_{0z}W_{1z}+(6\theta_{12}+2\theta_2)V_{0z}W_{0z}W_{1z}\\
&+(2\theta_{14}+2\theta_{15})V_0W_1V_{0zz}+(\theta_7+\theta_8)V_0V_{0z}V_{0zz}+2\theta_{12}W_1W_{0z}V_{0zz}+\theta_5V_{0z}W_{0z}V_{0zz}\\
&+(4\theta_3+4\theta_4)V_0V_1W_{0zz}+2\theta_5W_1^2W_{0zz}+2\theta_{12}W_1V_{0z}W_{0zz}+\frac{1}{2}\theta_5V_{0z}^2W_{0zz}\\
&+4\theta_2V_1W_{0z}W_{0zz}+2\theta_2V_0W_{1z}W_{0zz}+\theta_1W_{0z}W_{1z}W_{0zz}+(\theta_3+\theta_4)V_0^2W_{1zz}\\
&+2\theta_2V_0W_{0z}W_{1zz}+\frac{1}{2}\theta_1W_{0z}^2W_{1zz},\\
H_3=&(4\theta_1+8\theta_2+4\theta_3)V_0^2V_1+(6\theta_5-2\theta_6+4\theta_8)V_0W_1^2+(8\theta_{12}+8\theta_{15})W_1V_{0z}\\
&+(\frac{5}{2}\theta_5+\frac{1}{2}\theta_6+3\theta_8)V_0V_{0z}^2+(8\theta_2+8\theta_4)V_0V_1W_{0z}+(6\theta_7-2\theta_8)W_1^2W_{0z}\\
&+8\theta_{14}W_1V_{0z}W_{0z}+(\frac{5}{2}\theta_7+\frac{1}{2}\theta_8)V_{0z}^2W_{0z}+4\theta_3V_1W_{0z}^2+(\theta_{12}+\theta_{13}\\
&+2\theta_{15}+\theta_2+3\theta_4)V_0^2W_{1z}+(2\theta_{14}+2\theta_{15}+2\theta_3+2\theta_4)V_0W_{0z}W_{1z}\\
&+(\theta_{12}+\theta_2)W_{0z}^2W_{1z}+(\frac{1}{2}\theta_5+\frac{1}{2}\theta_6+\theta_8)V_0^2V_{0zz}+(\theta_7+\theta_8)V_0W_{0z}V_{0zz}\\
\end{aligned}
$$
$$
\begin{aligned}
&+\frac{1}{2}\theta_5W_{0z}^2V_{0zz}+(2\theta_{14}+2\theta_{15})V_0W_1W_{0zz}+(\theta_7+\theta_8)V_0V_{0z}W_{0zz}\\
&+2\theta_{12}W_1W_{0z}W_{0zz}+\theta_5V_{0z}W_{0z}W_{0zz},\\
H_4=&(\frac{1}{6}\theta_1+\frac{2}{3}\theta_2+\frac{1}{2}\theta_3)V_0^3+(\frac{1}{2}\theta_2+\frac{3}{2}\theta_4)V_0^2W_{0z}+(\frac{1}{2}\theta_3+\frac{1}{2}\theta_4)V_0W_{0z}^2+\frac{1}{6}\theta_2W_{0z}^3,\\
H_5=&(3\eta_1+5\eta_2)V_0V_1+2\eta_4W_1^2+2\eta_7W_1V_{0z}+\frac{1}{2}\eta_4V_{0z}^2+(3\eta_2+\eta_3)V_1W_{0z}\\
&+(\eta_2+\eta_3)V_0W_{1z}+\eta_2W_{0z}W_{1z},\\
H_6=&(\theta_5+\theta_6+2\theta_8)V_0^2W_1+(\frac{1}{2}\theta_{12}+\frac{1}{2}\theta_{13}+\theta_{15})V_0^2V_{0z}+(2\theta_7+2\theta_8)V_0W_1W_{0z}\\
&+(\theta_{14}+\theta_{15})V_0V_{0z}W_{0z}+\theta_5W_1W_{0z}^2+\frac{1}{2}\theta_{12}V_{0z}W_{0z}^2,\\
H_7=&(6\eta_4+2\eta_5)V_1W_1+(4\eta_4+4\eta_5)V_0W_2+(3\eta_7+\eta_8)V_1V_{0z}+(\eta_7+\eta_8)V_0V_{1z}\\
&+4\eta_4W_2W_{0z}+\eta_7V_{1z}W_{0z}+2\eta_4W_1W_{1z}+\eta_7V_{0z}W_{1z}.
\end{aligned}
$$

%%%%%%%%%%%%%%%%%%%%%%%%%%%%%%%%%%%%%%%%%%%%%%%%%%%%%%%%%%%%%%%%%%%%%%%%%%%%%%%%%%%%%%%%%%%%%%%%%%%%%%%%%%%

\end{document}